\documentclass[preprint2]{aastex}
\def\clh{\mbox{Cl$_6^{\rm hr}$}}
\def\cll{\mbox{Cl$_{\rm all}$}}
\def\costha{\mbox{$\cos\theta_{1/2}$}}
\def\costh{\mbox{$\cos\theta$}}
\def\vll{\mbox{$V_{\rm all}$}}
\def\vhr{\mbox{$V_{50}^{\rm hr}$}}

\def\kpch{\mbox{$h^{-1}$kpc}}
\def\lcdm{{$\Lambda$CDM}}
\def\mpch{\mbox{$h^{-1}$Mpc}}

\def\msun{\mbox{M$_\odot$}}
\def\msunh{\mbox{$h^{-1}$M$_\odot$}}
\def\rhoh{\mbox{h$^2$M$_{\odot}$Mpc$^{-3}$}}
\def\mv{\mbox{$M_{\rm vir}$}}
\def\mh{\mbox{$M_{h}$}}

\def\ome{\mbox{$\Omega_0$}}
\def\omel{\mbox{$\Omega_\Lambda$}}

\def\rh{\mbox{$R_{h}$}}
\def\rv{\mbox{$R_{\rm vir}$}}
\def\rt{\mbox{$R_{\rm tr}$}}

\def\teta{\mbox{$\theta_{1/2}$}}
\def\eg{e.g.,}
\def\nbody{\mbox{$N-$body}}
\def\vm{\mbox{$V_{\rm max}$}}
\def\cnfw{\mbox{c$_{\rm NFW}$}}
\def\cq{\mbox{c$_{\rm 1/5}$}}
\def\cd{\mbox{c$_{\delta}$}}
\def\sigt{\mbox{$\sigma_t$}}
\def\sigc{\mbox{$\sigma_c$}}

\received{2005 March 15}
\begin{document}
\shorttitle{Dependence on environment of CDM halo properties}
\shortauthors{Avila-Reese et al.}

\title{The dependence on environment of Cold Dark Matter Halo properties}

\author{Vladimir Avila-Reese\footnote{Instituto de Astronom\'{\i}a, U.N.A.M., A.P. 70-264, 04510, M\'exico, D.F., M\'exico} ,  Pedro Col\'{\i}n\footnote{Centro de Radioastronom\'{\i}a y Astrof\'{\i}sica, U.N.A.M., A.P. 72-3 (Xangari), Morelia, Michoac\'an 58089, M\'exico} ,  Stefan Gottl\"ober\footnote{Astrophysical Institute Potsdam, An der Sternwarte 16, 14482 Potsdam, Germany},} 

\author{Claudio Firmani\footnote{Osservatorio Astronomico di Brera, via E.Bianchi 46, I-23807 Merate, Italy} $^{,1}$, and Christian Maulbetsch$^3$}


\begin{abstract}
A series of high-resolution $\Lambda$CDM cosmological N-body 
simulations are used to study the properties of {\it galaxy-size} dark halos 
as a function of global environment. We analyse halos in three types of 
environment: ``cluster'' (cluster halos and their surroundings), 
``void'' (large regions with density contrasts $\lesssim -0.85$), and ``field'' 
(halos not contained within larger halos). 
We find that halos in clusters have a median spin parameter $\sim 1.3$ times 
lower, a minor-to-major axial ratio $\sim 1.2$ times lower (more spherical), 
and a less aligned internal angular momentum than halos in voids and
the field. For masses $\lesssim 5\times 10^{11}\msunh$, halos in cluster regions 
are on average $\sim 30-40\%$ more concentrated and have $\sim 2$ times higher 
central densities than halos in voids. While for halos in cluster regions the 
concentration parameters decrease on average with mass with a slope of $\sim 0.1$, 
for halos in voids these concentrations do not seem to change with mass.
When comparing only parent halos from the samples, the differences are less 
pronounced but they are still significant. We obtain also the maximum circular 
velocity-- and rms velocity--mass relations. These relations are shallower 
and more scattered for halos in clusters than in voids, and for a given circular 
velocity or rms velocity, the mass is smaller at $z=1$ than at $z=0$ for all 
environments. At $z=1$, the differences in the halo properties with environment 
almost dissapear, suggesting this that the differences were stablished
mainly after $z\sim 1$. The halos in the cluster regions undergo more
dramatic changes than those in the field or the voids. The differences in 
halo properties with environment are owing to (i) the dependence of 
halo formation time on global environment, and (ii) local effects as tidal 
stripping and the tumultuos histories that halos suffer in high-density regions. 

We calculate seminumerical models of disk galaxy evolution using halos
with the concentrations and spin parameters found for the 
different environments. For a given disk mass, the galaxy 
disks have higher surface density, larger maximum circular velocity
and secular bulge-to-disk ratio, lower gas fraction, and are redder as
one goes from cluster to void environments. Although all these trends agree 
with observations, the latter tend to show more differences, suggesting this 
that physical ingredients not considered here as missalignment of angular momentum,
halo triaxility, merging, ram pressure stripping, harassment, etc. play an 
important role for galaxy evolution, specially in high-density environments.

\end{abstract}
 
\section{Introduction}

The understanding of formation and evolution of cosmic structures in the 
Universe is a fundamental problem in astrophysics as well as an 
observational testbed for cosmological, particle and high-energy physical 
theories. The current paradigm in this topic is the inflationary motivated
$\Lambda-$Cold Dark Matter ($\Lambda$CDM) hierarchical scenario. 
According to this scenario, the backbone of galaxy formation and 
evolution are the CDM halos, which emerge from a Gaussian primordial density 
fluctuation
field and assemble through gravitational processes. The $\Lambda$CDM scenario 
succesfully reproduces most of cosmological observational data,
from the angular anisotropies of Cosmic Microwave Radiation to galaxy clustering 
\citep[see for a review][]{Frenk02}. The confrontation of models and observations 
at the level of galaxies is more difficult due to the complexity of the baryonic 
processes such as gas cooling, hydrodynamics, star formation (hereafter SF), and
feedback. Several approaches have been introduced to model these processes in the 
context of the $\Lambda$CDM scenario and their results are encouraging, however, 
some difficulties still remain \citep[see for a review][]{FAR03}.  To gain 
insight on galaxy formation 
and evolution, as well as on the galaxy observables that can be used directly
to test the $\Lambda$CDM scenario, one should identify which galaxy properties 
are clean tracers of CDM halo properties. The comparison of the 
dependences on global environment of both the simulated CDM halo and the observable 
galaxy properties may shed ligth on this undertaking.

\subsection{Observational dependences of galaxy properties on environment}

It is well known that several galaxy properties vary as a function of 
environment. The morphologycal type of galaxies is earlier in the locally 
denser regions (morphology-density relation), being the fraction of ellipticals 
maximal in cluster cores \citep{Dressler80} and enhanced in rich \citep{PG84} 
and poor \citep{Tran02, Tov04} groups. An even tighter 
correlation has been found between morphology and cluster- or group-centric radius 
\citep{Whitmore93, Dom02}.  The extension of the 
morphology-density and morphology-radius relations to low local-density environment 
(cluster outskirts, low mass groups, field) has been a matter of debate 
\citep[\eg][]{PG84, Dom02}. From an analysis 
of SDSS data, \citet{Goto03} have found that (i) in the sparsest regions 
both relations flatten out, (ii) in the intermediate density regions (e.g., 
cluster outskirts) the intermediate-type galaxy (mostly S0s) fraction 
increases towards 
denser regions whereas the late-type galaxy fraction decreases, and (iii) in 
the densest regions intermediate-type fraction decreases radically and early-type 
fraction increases. In a similar way, an study based on 2dF data of the luminosity 
functions in clusters and voids show that the population of faint 
late-type galaxies dominates in the latter, while, in contrast, very bright 
early-late galaxies are relatively overabundant in the former \citep{Croton05}. 
These an other studies \citep[\eg][]{Dom01} suggest 
that the origin of the morphology-density (or morphology-radius) relation could 
be a combination of (i) {\it initial cosmological conditions} and (ii) of 
{\it external mechanisms} (ram-pressure and tidal stripping, thermal evaporation 
of the disk gas, strangulation, galaxy harashment, truncated SF, etc.) that operate 
mostly in dense environments, where namely the relation steepens significantly.

The morphology-environment relation evolves. It systematically flattens with $z$
in the sense that the grow of the early-type (E+S0) galaxy fraction with density 
becomes less rapid \citep{Treu03, sm04, Goto04,Postman05}, 
the main change being in the high-density population fraction. Postman et al. 
conclude that the observed flattening of the relation up to $z\sim 1$ is due 
mainly to a deficit of S0 galaxies and an excess of Sp+Irr galaxies relative to 
the local galaxy population; the E fraction-density relation does not appear
to evolve over the range $0<z<1.3$.

Observational studies show that other properties besides morphology vary 
with environment locally \citep[\eg][]{Dressler85, GH85, 
ZF93}. More recently it was shown that the galaxy properties 
most sensitive to environment are the integral colour and specific SF 
rate \citep[\eg][]{Ba98,Ba04, Ko01, Tran01, Pimbblet02, Gomez03, 
Lewis02, Kauffmann04, Hogg04, Tanaka04}, the steepest correlation being for 
intermediate mass galaxies \citep{Kauffmann04} or faint galaxies
\citep{Tanaka04}. The dependences of both properties on environment extend 
typically to lower local densities than the dependence for morphology. 
These properties are tightly related to the galaxy SF history, 
which in turn depends on internal formation/evolution processes related 
directly to initial cosmological conditions \citep[\eg][]{KWG93,
Baugh96,Somerville99,AF00}
as well as to external astrophysical mechanisms able to inhibit or 
induce SF activity. It is difficult to disantangle the role of these 
internal or external factors, in particular in the densest environments. An
analysis of the influence of environment on the CDM halo properties and their
evolution certainly would help to understand the role of initial conditions.

There are also some pieces of evidence that the scale lenghts of spirals in 
clusters are systematically smaller than those of spirals in the field 
\citep{Agerri04}. Besides, it was shown that the fraction of low 
surface brightness (LSB), blue galaxies increases toward low-density 
environments \citep[\eg][]{Bothun93, Rojas04}, mainly in the outskirts 
of filamentes and in voids \citep{RB04}.  The 
LSB galaxies may have formed in halos with high and well-aligned angular momentum 
\citep[\eg][]{Dalcanton97, FA00, AF00, Boissier03} and/or have had a low SF 
rate history in comparison with high surface brightness galaxies due 
to their low-density environment. To explore which one of these processes 
dominate it would be helpful to know if there are any differences in the halo 
angular momentum distribution and its alignment between voids and other environments.

\subsection{Theoretical results and expectations}

From the analysis of a $60\mpch$ box simulation with a particle mass of 
$1.57 \times 10^{9}\msun$, \citet{AR99} found that the outer density profiles 
of galaxy-size halos in cluster cores decline sharper and their
concentrations are higher than for ``field'' halos
\citep[see also][]{Ghigna98, OH99, Klypin99, Bullock2001b}. 
The last authors explored also the differences in concentration between
distinct or parent halos (halos not contained within larger ones) and subhalos
(halos contained inside larger halos), with the result that the latter
have typically larger concentrations than the former. As suggested by
the referee, one should differentiate the effect of the ``local'' 
environment, which is simply due to a halo being subhalo, from the
global (large-scale) environment.
Some dependence of the halo mass-circular velocity relation on global
environment was also reported by \citet{AR99}. 

\citet{LK99} explored how halo mass function, 
formation redshift, concentration, shape and spin parameter $\lambda$ 
change with the density contrast of the local halo environment. The particle 
mass in their simulations was $2 \times 10^{10}\msun$ or more. They found that 
only the mass function varies with environment, suggesting that any 
dependence of observable galaxy properties on environment can be established 
because only the halo mass influences on these properties. \citet{AD02} used 
constrained initial conditions in order to
produce in the same box two extreme environments, a void and two
clusters (the particle mass of their simulation was $4.2 \times 10^{9}\msun$). 
They found some differences in the distribution of the spin parameter and 
the mass-velocity dispersion relation between cluster and void environments. 
 
A general conclusion of all previous numerical works is that present-day 
CDM halo properties and correlations do not strongly change with global 
environment. However, as mentioned above, some influence of environment 
on halo properties and correlations has been reported. Halos in dense 
environments, as clusters, are expected to collapse earlier 
than halos in less dense environments. Besides, halos in 
the high density environments suffer tidal stripping and frequent violent 
mergers. It is still an open issue the extent of the influence of 
environment on halo properties as well as its potential
effects on the baryonic galaxies which form within these halos.

In this paper we will analyze high-resolution simulations of extreme
environments, namely selected cluster and void regions, and
compare the properties and correlations of the galaxy-size CDM halos 
from these simulations, both at redshift $z=0$ and $z=1$. Our aim is to 
explore to what extent the CDM halo properties that affect observable 
galaxy properties change with the global environments. We will analyze
also distinct galaxy-size halos (``field'' halos) and compare
them with those in the cluster and void simulations. 

Seminumerical and semianalytical models have shown that most of 
the properties and correlations of galaxies formed within CDM 
halos depend on (i) the halo mass aggregation history (MAH) and its 
dominating regime (accretion or merging) as well as on the halo concentration 
that is determined mainly by the MAH (in particular the typical formation
time) for a given mass, (ii) the halo spin 
parameter and angular momentum distribution, and (iii) the halo mass function 
\citep[\eg][]{KWG93, Baugh96, MMW98, AF00, FA00, vdBosch00}. 
The dependence on environment of the halo mass function has been 
extensively analyzed in numerical works \citep{LK99, Stefan03,Mo2004}. 
Our comparative study here
will be focused on some halo structural and angular momentum parameters:
concentration, central density, shape, spin parameter, and the angular 
momentum internal alignment. Elsewhere we will present results related to halo 
evolution (MAH) depending on environment \citep[see also][]{Gardner01,Gott01}.

In \S 2 we describe the method and simulations carried out and present the halo 
samples to be analyzed. In \S\S 3.1 several concentration parameters 
as well as the central density of halos from the different environments (samples)
are presented and disscused for $z=0$ and $z=1$. The ellipticity
distribution function of halos for the different samples 
is computed in \S\S 3.2. Section 4 is devoted to the
analysis of the spin parameter and internal angular momentum alignment of halos
as a function of environment. In \S 5 we present and discuss the maximum
circular velocity-- and velocity dispersion--mass relations of halos
in the different environments. In \S 6 we discuss 
the implications that the differences in the halo properties as a function of 
environment can have on the properties of the disks formed inside these halos. 
A summary of the results and our main conclusions are given in \S 7.

\begin{deluxetable}{cccccc}
\tablecolumns{5}
\tablewidth{0pc}
\tablecaption{Parameters of simulations}
\label{tab:simu}
\tablehead{\colhead{simulation} & \colhead{Box} & \colhead{N$_{\rm part}$} &
\colhead{Mass resolution} & \colhead{Force resolution} & \colhead{$z_{\rm start}$} \\
& (\mpch) &     &(\msunh) & (\kpch) & }
\startdata
Full box  \\
80  & 80  & $128^3$    & $2.0 \times 10^{10}$ & 9.8 & 50 \\
60A  & 60  & $256^3$    & $1.1 \times 10^9$    & 2.0 & 35 \\
60B  & 60  & $256^3$    & $1.1 \times 10^9$    & 2.0 & 35 \\
50   & 50  & $128^3$    & $5.0 \times 10^9$    & 6.2 & 60 \\
Multiple mass  \\
\cll & 80  & $(512^3) $ & $3.2 \times 10^8$    & 2.4 & 50 \\
\clh & 80  & $(1024^3)$ & $4.0 \times 10^7$    & 0.6 & 50 \\
\vll & 80  & $(1024^3)$ & $4.0 \times 10^7$    & 1.2 & 50 \\
\vhr & 50  & $(1024^3)$ & $2.5 \times 10^7$    & 0.8 & 60 \\
\enddata
\end{deluxetable}

\section{Method and simulations}

In our numerical simulations we adopt the flat cosmological model with
a non-vanishing cosmological constant (\lcdm) using the following
cosmological parameters: $\Omega_0=0.3$, $\Omega_{\Lambda}=0.7$, and
$h=0.7$.  The matter power spectrum is normalized to $\sigma_8=0.9$,
where $\sigma_8$ is the present linear rms amplitude of mass
fluctuations in spheres of radius 8\mpch.

The simulations were run with the Adaptive Refinement Tree (ART) code
\citep{KKK97}.  The ART code achieves high spatial resolution by
refining the base uniform grid in all high-density regions with an
automated refinement algorithm. Initial conditions have been
calculated either with the transfer function given by \citet{KH97}
(boxes of 60 and 80 \mpch) or using the numerical results 
of the Boltzmann code 
kindly provided by W. Hu. Up to a few percent both
transfer functions coincide and at the same level they coincide also
with the transfer function provided by CMBfast. For the multiple mass
simulations we have used the mass refinement technique described by
\citet{Klypin01}.

The halos are identified by the Bound Density Maxima (BDM) algorithm
\citep{KH97,Klypin99}.  The BDM algorithm first finds positions of
local maxima in the density field.  Once centers of potential halos
are found, the algorithm identifies halos around them and removes
particles which are not bound to those halos.  This procedure also
detects subhalos of larger objects -- halos inside halos (for example,
satellites of galaxies or galaxies in clusters).  Particles of a
subhalo are bound to both the subhalo and to the larger halo.

Our study is focused on the analysis of the properties of galaxy-size
CDM halos formed in two extreme environments: the high density environment of
clusters and the low density environment of voids. To this end we have
at first performed simulations with low mass resolution.  However,
already for this simulation the initial displacements and velocities
of the particles were calculated using all waves ranging from the
fundamental mode $k=2\pi/L$ to the Nyquist frequency $k_{\rm
 Ny}=2\pi/L\times N_1/2$, where $N_1=1024$ corresponds to the maximum
possible mass resolution (cf. Table 1). To get the initial conditions
of the low resolution run we merge $8^3$ particles to a more massive
one. After running an inexpensive simulation with the resulting
$128^3$ particles we can select regions of interest. Then we use the
original sample of small-mass particles to resimulate the regions of
interest using the multi-mass technique described in \citet{Klypin01}.

As high density regions we have selected spheres centered on 14
clusters. The radius of these spheres was typically about three virial
radii of the cluster. Thus a significant fraction of resimulated halos
are not within the virial radius of one of the clusters, but these
halos are in the high-density environment around clusters. Voids have
been selected as described in \citet{Stefan03}. The radius of
resimulation area centered on voids has been taken typically about
10\% smaller than the estimated void radius to be sure that no
massive objects around the low density region will enter the region
with high mass and force resolution, otherwise all the integration
time would be allocated to such a massive object. We have performed
high resolution simulations of selected regions in a 80 $\mpch$ box
and in a 50 $\mpch$ box (simulations 80 and 50, respectively). For 
comparison purposes, we have also
analyzed a sample of galaxy-size ``field'' halos from two 60\mpch\
box simulations with the same cosmological parameters as simulation
80 (simulations 60A and 60B).  The parameters of
the simulations are summarized in Table 1. The brackets around the
particle numbers denote that all low mass particles were merged
outside the regions of interest.

The ``field'' halos in the 60\mpch\ box are those called distinct or 
parent, i.e. halos which are not contained inside larger halos 
\citep[][]{AR99,Bullock2001b}. Most of the parent halos
are truly isolated ($\sim 80\%$); the rest are mostly 'binary' halos,
i.e. with companions of masses larger than 1/3 the mass and within
distances up to 3 times the radius of the given halo.

\subsection{The halo samples according to the environment}
  
From the simulations described above, we have five independent subsamples of
halos (see Table 2):  two cluster subsamples selected from 
simulation 80, one made up of halos from the very high-resolution run \clh, 
and the other built from the sum of the halos belonging to the 14 clusters,
\cll; both subsamples constitute the CLUSTER sample, where $\sim 55\%$ of 
the galaxy-size halos are parent (those outside the clusters), and 
$\sim 45\%$ are subhalos (mostly those within the cluster virial radii); two void 
subsamples, one 
consisting of halos from the simulation \vhr, and the other built of halos 
from the 5 voids resimulated from the 80 simulation, \vll; boths subsamples 
constitute the VOID sample, where more than $90\%$ of the 
galaxy-size halos are parent ones; one FIELD halo sample composed of two
subsamples of galaxy-size parent halos from both the 60A and 60B simulations.

Column (2) of Table 2 gives the number of regions (clusters or voids)
in each (sub)sample resimulated with high resolution; this does not apply 
for the FIELD sample, where all the parent galaxy-size halos
in the boxes are taken into account.  The number of halos with more than 
500 particles in each one of the samples is given in column (3). We set an 
upper limit mass of $3 \ 10^{13}\msunh$ for that we define as a galaxy-size 
halo. In column (4) we give an estimate of the density contrast 
[$(\bar{\rho}_{\rm region}-\rho_u)/\rho_u$] of our different environments.
For the samples where several regions were used (\cll\ and \vll),
the mean of the density contrasts and the standard deviation are reported;
for the  ``field'' halos, the density contrast is around zero. 
The mean halo mass ($N_p>500$ particles) and the median spin parameter
of each sample are given in columns (5) and (6), respectively.

Finally, we note that in multimass simulations we discard the halos 
contaminated by massive
particles if their fraction is larger than 5\% of the total halo mass.
Also fake halos (recognized by a too low central density or 
flat or decreasing density profiles toward the center)
have been removed; the criteria to remove these halos are 
a density in the central bin below $\sim 5\
10^{15}$\rhoh\ and/or similar or increasing with radius densities
in the the first 5 radial bins.
To study the spin parameter distribution we demand a halo to have a
minimum of 500 particles. For concentration and central density
determinations we demand more than 2500 particles.

\begin{deluxetable}{cccccc}
\tablecolumns{5}
\tablewidth{0pc}
\tablecaption{Samples of halos}
\label{tab:sample}
\tablehead{\colhead{sample} & \colhead{$N_{\rm regions}$} & \colhead{$n_h(>500)$} & \colhead{contrast} &
\colhead{$<M_h>$} & \colhead{$\lambda'_{med}$} \\
 &   &   &   &  (\msunh) & }
\startdata
\cll\ & 14 &  421 & $37.8\pm 18.2$ & $5.3 \times 10^{11}$ & 0.024 \\
\clh\ &  1 &  318 & 44.7 & $1.0 \times 10^{11}$ & 0.024 \\
\vll\ &  5 &  333 & $-0.85\pm 0.05$ & $4.9 \times 10^{10}$ & 0.030 \\
\vhr\ &  1 &  249 & -0.85  & $2.2 \times 10^{10}$ & 0.033 \\
FIELD & .. & 2990 & $\sim 0$  & $2.2 \times 10^{12}$ & 0.033 \\
\enddata
\end{deluxetable}

\section{Structural properties}

The virialized CDM halos present a diversity of spherically
averaged density profiles \citep{AR98, AR99, Jing99,TKGK04} that 
cluster around the Navarro, Frenk \& White (1997; 
hereafter NFW) fit. High-resolution simulations show that in the 
center the density profile is typically steeper than the NFW profile 
\citep{Moore99, Klypin01, Power03, Navarro04}. Unfortunately, 
there is not still a clear theoretical understanging
of the origin of the CDM halo density profiles \citep[see for some
interesting results][and more references therein]{Manrique03}.
Several works have shown that the shape of the profiles or their 
concentration parameter if they are fitted 
by a NFW profile --which is not always a good fit, depend
on the halo MAH \citep{AR98, FA00, Risa02} or even on the cosmological
initial conditions \citep{AR03}. It was shown in \citet{AR99} that 
the outer density profile of CDM halos change with environment.
Following, we will explore whether or not the halo concentrations, central 
density, and internal shape alignment change with environment.

\subsection{Concentrations and inner density}

We construct spherically averaged density profiles for all the halos
studied here. We can roughly characterize the profiles by a concentration
parameter only, because the profiles have approximately similar shapes. The 
use of one parameter allows us to attain an easy statistical comparison between 
VOID, CLUSTER and FIELD samples. In the CDM hierarchical scenario one 
expects that the concentration of halos will be typically higher for less 
massive halos, because they assemble most of their mass earlier than larger 
mass halos (e.g., NFW). 

For the NFW profile, the concentration is defined as
$\cnfw \equiv \rv/r_s$, where \rv\ is the virial radius and $r_s$ 
the radius where $d$ln$\rho/d$ln$r=-2$ (NFW). The virial radius \rv~ is 
defined as the radius where the average halo density is $\Delta$ times 
the background density according to the spherical top-hat model.
The parameter $\Delta$ depends on
epoch and cosmological parameters (\ome,\omel); for a flat \lcdm\
model, $\Delta \sim 337$ and 203 at $z = 0$ and $z = 1$, respectively. 
As mentioned above, not all halos can be well fitted by a NFW profile. 
One may introduce a concentration parameter that does not depend on a 
particular fitting profile, for example the \cq\ concentration
defined as $\cq \equiv \rv/r_{1/5}$, where $r_{1/5}$ is the radius
where 1/5 of the virial mass is contained.
  
For both the \cnfw\ and \cq\ concentrations one needs to measure \rv. On 
one hand, this is not possible for a large fraction of subhalos because 
they are typically truncated at a radius $\rt < \rv$. 
On the other hand, both \cnfw\ and \cq\ trace the global halo concentration, 
while galaxies form only in the inner parts
of the halos.  Thus, it is desirable to introduce a concentration parameter 
whose definition is independent of the virial radius, and which measures 
concentration in the inner/intermediate halo regions. We introduce here the 
concentration parameter \cd. To calculate \cd, we find the radii within which 
the halo average density is equal to $4\times 10^4$ and $4\times 10^3$ times 
the background density, respectively, and compute the enclosed mass within
these radii. The ratio of these masses is \cd. For NFW profiles in the range 
of galaxy-size halos, the typical radii where the halo overdensities become 
$4\times 10^4$ and $4\times 10^3$ are in the range of $\sim 0.10-0.45$\rv, 
respectively, tracing therefore the inner to intermediate regions. 
For example, for \cnfw=12, these radii are 0.11 and 0.35\rv, respectively, 
and the ratio of the masses enclosed within these radii is \cd=0.33.  
We define formally the truncation radius \rt\ as the radius 
where the outer density profile begins to systematically flatten out or 
increase (d$\rho(r)$/d$t>0$). The radius of the halo, \rh\, is defined 
as the minimum beetwen \rv\ and \rt.  The mass enclosed within \rh\ is 
the halo mass \mh.

According to our resolution tests, halos with more than $\sim 2500$ 
particles are needed in order to estimate reliable concentration parameters.
This leaves us with halos more massive than $\sim 10^{11}$\msunh\ 
and $\sim 8 \times 10^{11}$\msunh\ in the void \vll\ and
cluster \cll\ samples, respectively. Since the voids are
defined in such a way that they contain only halos less massive than 
$\sim 4 \times 10^{11}$\msunh\ \citep{Stefan03}, the ranges of
masses in both samples do not overlap and our comparisons should 
be based on extrapolations of the mass-concentration dependences. 
However, we have one cluster resimulated with the same resolution
as the voids, \clh. Therefore, in this simulation the halos resolved 
with more than 2500 particles have masses larger than $\sim 10^{11}$\msunh. 
There is also a void simulation, \vhr, for which halos resolved with more than 2500 
particles have masses larger than $\sim 5\ 10^{10}$\msunh, allowing us to expand 
the range of masses of void halos. 

\subsubsection{Results at z=0}

Upper panel of Fig. 1 shows \cnfw\ vs \mv\ for the VOID (circles), 
\cll\ (crosses), \clh (squeletal triangles) and FIELD (dots) 
{\it non-truncated} halos with an acceptable NFW profile fit. 
Non-truncated halos are those that 
attain the virial radius without any systematical flattening 
or increasing of the outer density profile. They are mostly parent halos. 
The quality of the NFW fit is evaluated roughly through 
the variance of the fit divided by the weighted average of the individual 
variances of the data (each radial bin is assigned a weight according 
to the number of particles inside it). Since the fit is in logarithmic 
variables, the root square of 
this quantitiy can be interpreted as a measure of the logarithmic difference 
between the model fit and the data. The latter can be expressed as a 
percentage deviation; from visual inspections, we have 
found that an acceptable NFW fit to the density profile is when this deviation 
is less than $\sim 15\%$. More than 80\% of the non-truncated halos obey this 
condition.  The parent halos in the CLUSTER sample are highlighted with an open
square. As expected, most of the non-truncated halos are parent ones.
In the VOID sample, more than 95\% of the halos with more than 
2500 particles are parent ones, then, we do not highlight these halos
in the figures in order to avoid overplotting. The FIELD sample has been 
constructed only with parent halos.

In the lower panel of Fig. 1, we show \cq\ vs. \mv\ using the same
symbols than in the upper panel. We recall that the definition of \cq\ is 
not related to any postulated density profile. Therefore, in the lower
panel halos with non-acceptable NFW fits are also included. In Fig. 2
we present \cd\ vs \mh\ (upper panel), and $\rho_{-2}$ vs \mh\ (lower panel) 
with the same symbol code as in Fig. 1. The parameter $\rho_{-2}$ is a 
characteristic inner density measured at the radius where the slope of the 
logarithmic density profiles becomes $-2$. The definitions of the concentration
\cd\ and of $\rho_{-2}$ do not depend on \rv, therefore, 
in Fig. 2 both truncated and non-truncated halos are included. 


\begin{figure}[htb!]
\vspace{11.5cm}
\includegraphics{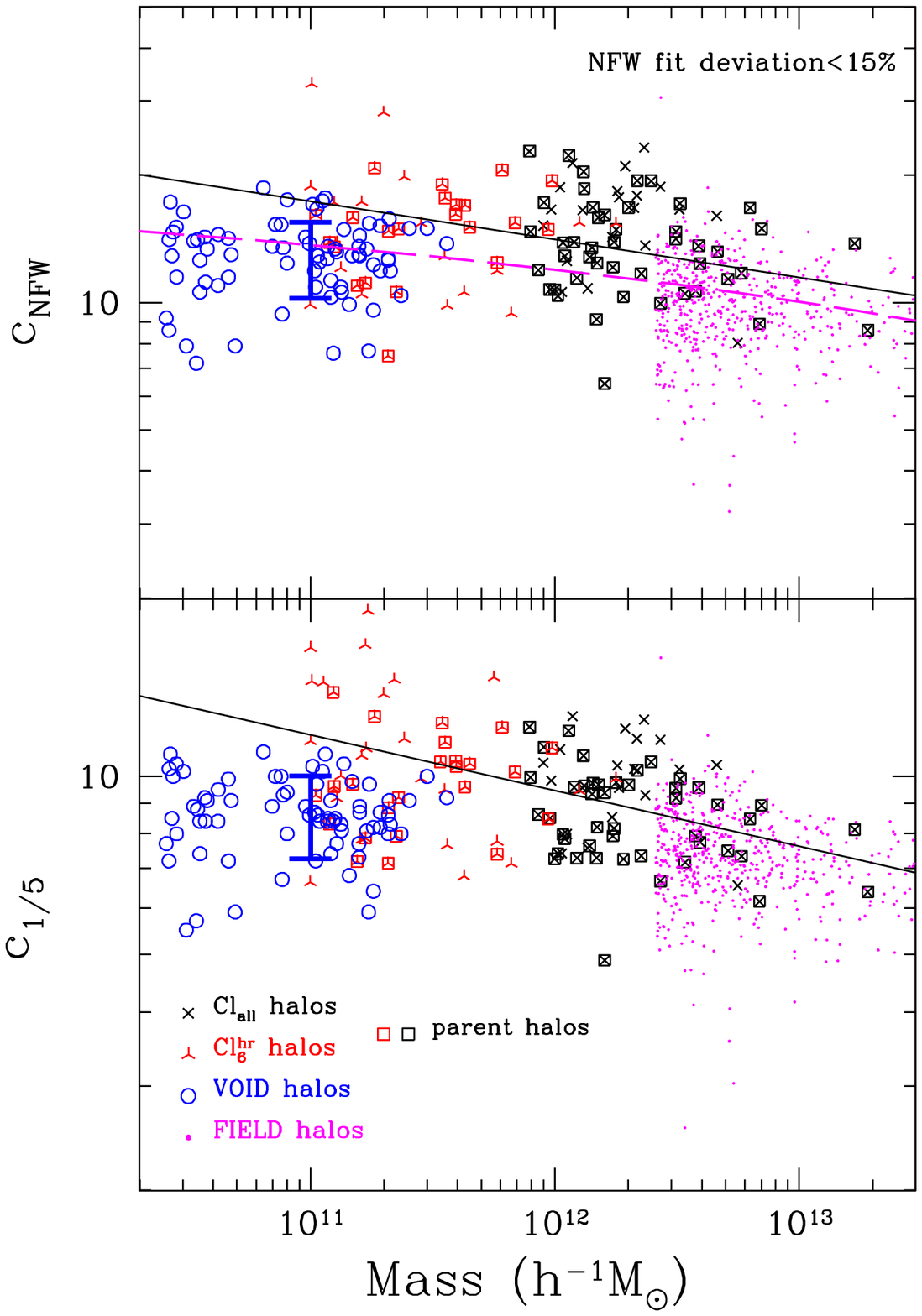}
\caption{The NFW and 1/5 concentration parameters vs. virial mass \mv
for the \cll, \clh, VOID (\vll + \vhr), and FIELD samples.
The symbols are explained in the lower panel. Halos with truncated
radius are not included in the plots. In the upper panel were included 
only halos whose density profiles are fitted well by the NFW model. Solid lines
are direct linear regressions to the CLUSTER (\cll + \clh) sample, while 
the error bars
are the dispersions of the VOID sample (the abscise is arbitrary since
there is no dependence on mass for this sample). Thick dashed line in 
the upper panel is the prediction of a toy model calibrated to N-body 
simulations of relaxed isolated halos \citep{ENS2001}.}
\end{figure}


\begin{figure}[htb!]
\vspace{11.5cm}
\includegraphics{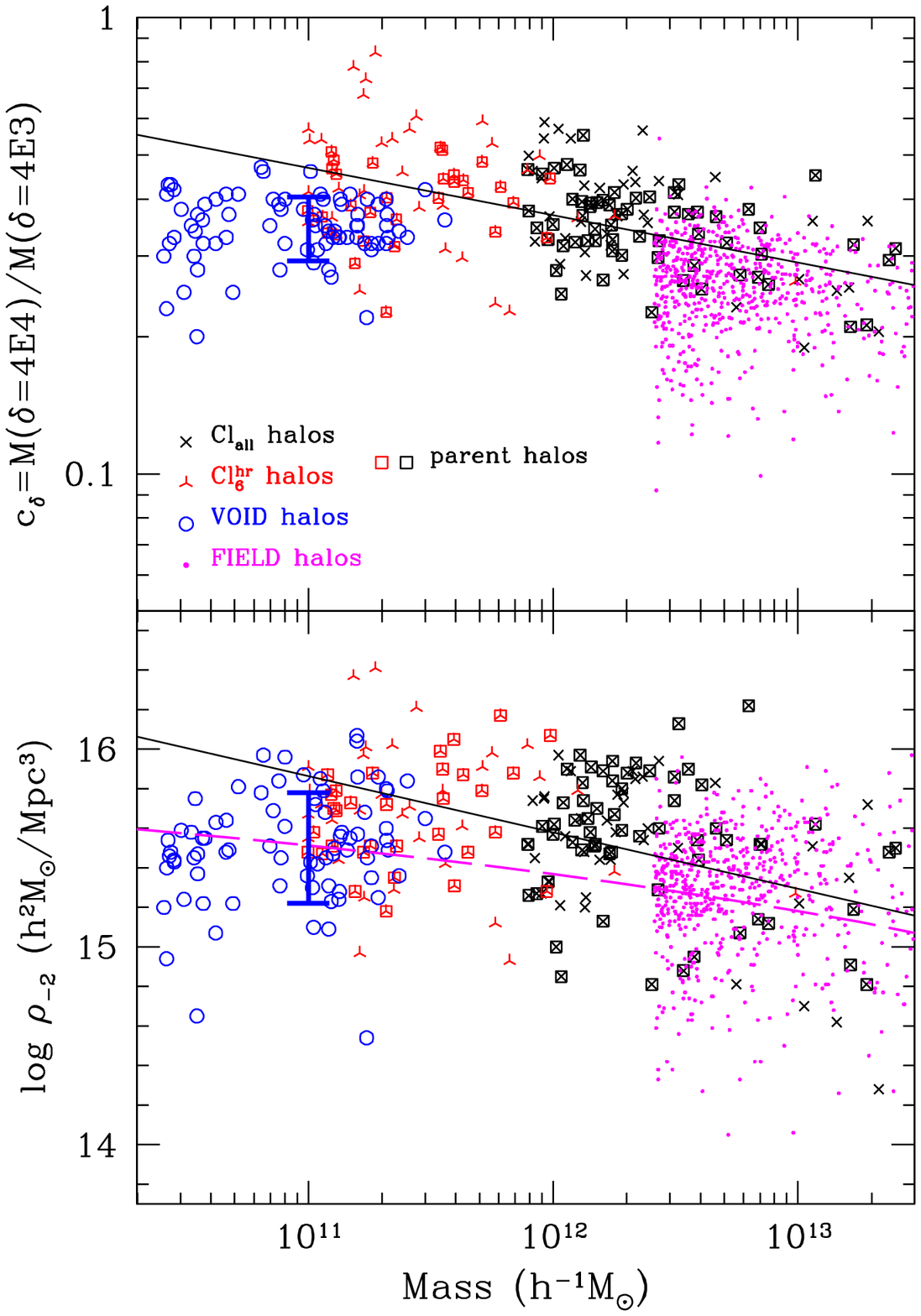}
\caption{Inner concentration parameter, \cd, and $\rho_{-2}$ density vs. halo mass \mh\
for the \cll, \clh, VOID (\vll + \vhr), and FIELD samples.
The symbols are explained in the upper panel.  Solid lines
are direct linear regressions to the CLUSTER sample, while the error bars are
the dispersions of the VOID sample (the abscise is arbitrary since
there is no dependences on mass for this sample). Thick dashed line in 
the lower panel is the prediction of a toy model for the NFW profile 
calibrated to N-body simulations of relaxed isolated halos \citep{ENS2001}.}
\end{figure}


The concentration parameters and the central density systematically
decrease with mass for the CLUSTER sample. In all the cases the scatter
is large and the correlations are weak. The Pearson correlation 
coefficients of the linear regressions (in logarithmic variables) are 
between -0.35 and -0.55. As a reference, in Table 3 the parameters of the
direct linear regressions are given. 
We also report the global dispersion of each linear regression, $\sigma$.
If the scatter is Gaussian distributed (lognormal in linear variables)
and independent of mass, then $\sigma$ represents the 1$\sigma$ (68\%) 
variation of $\Delta$(log$Y$) at any mass, where $Y$ is one of the concentration 
parameters or $\rho_{-2}$. By multiplying $\sigma$ by 2.3, 
one get the width of the lognormal distribution (in natural
logarithm), $\sigma_{\rm ln}\approx 2.3\sigma$.

Solid lines in Figs. 1 and 2 indicate the 
linear regresions to the \cll+\clh (CLUSTER) sample data.  The slopes of the 
regressions for \cnfw\ and \cq\ are shallow. If we take into account all the 
halos, and not only those that are well fitted by the NFW profile, then the 
slope for $\log$\cnfw\ vs $\log$\mv\ is steeper, $-0.14$, but the scatter 
becomes larger (from $\sigma= 0.13$ to 0.17).  Note that a significant 
fraction of the halos in the cluster simulations ($\sim 55\%$) are outside 
the virial radii of the cluster halos, i.e. they are not subhalos, but 
are in any case in overdense regions.    

The \cd\ parameter is independent of the definition of the halo
radius (virial or truncation); in this sense, \cd\ is a more direct 
estimator of halo mean (inner) concentration, and it allows to include
almost all the subhalos in the sample.  As Fig. 2 and Table 3 show, 
for halos in CLUSTER sample, \cd\ on average decreases as \mh\ increases. 
This confirms the decreasing of the physical concentration of halos/subhalos 
in dense environments as their masses (virial or truncated) increase.
 
The dispersions of the linear regressions presented in Table 3 give an
estimation of the scatter of the concentration parameters and $\rho_{-2}$.
As \citet{Bullock2001b} found, the scatter for \cnfw\ does not depend 
significantly on mass and it has a lognormal distribution. These authors 
report that the 1$\sigma$ deviation of $\Delta$(log$\cnfw$) is 0.18 
for distinct halos, and 0.24 for subhalos
(the correct values are actually 0.14 and 0.18, respectively,
J. Bullock, private communication).
In the understanding that the dispersion $\sigma$ reported in Table 3
is similar to the 1$\sigma$ deviation of $\Delta$(log$\cnfw$) (see above),
we find that the scatter of $\cnfw$ for our prunned CLUSTER sample 
(including both parent halos and subhalos) is similar to that of the 
distinct halos at all environments in \citet{Bullock2001b}; by including 
halos with density profiles not well fitted by the NFW model 
\citep[as in][]{Bullock2001b}, the scatter of the CLUSTER sample increases 
to 0.17. By separating the prunned sample in parent halos and subhalos, 
the 1$\sigma$ variations of $\Delta$(log$\cnfw$) are 0.12 and 0.17, 
respectively. We have calculated also the value of the 1$-\sigma$ deviation 
of $\Delta$(log$\cnfw$) in a range of masses from $5 \ 10^{11} - 2 \ 10^{12}\msun$
for the halo/subhalo sample; the value is the same than that found through 
the linear regression to all the mass range, 0.13.

For the VOID (low mass) halo sample we do not find any trend of \cnfw, \cq,
\cd, and $\rho_{-2}$ with mass, although the range of masses studied 
is too small to claim for definitive conclusions.  It seems that the 
halo assembling process in low-density environments is so slow, constant
and late that a posible mass-formation epoch relation \citep{LC93, AR98} 
loses sense; the dependence of concentration on mass is attributed to this relation 
(e.g., NFW).  In the case of halos in VOID sample, almost all of them are parent ones.  
The error bars in the left sides of Figs. 1 and 2 represent the (logarithmic) 
1$\sigma$ dispersions of \cnfw, \cq, \cd\ and $\rho_{-2}$ for the VOID sample 
(see also Table 3 for roughly the same results). Since there is no dependence 
of these quantities on mass, the abscise of the bars is arbitrary. The scatters
of the concentrations and  $\rho_{-2}$ in the VOID halo sample are significantly
smaller than the ones in the CLUSTER sample. For example, the 1$\sigma$ 
variation of $\Delta$(log$\cnfw$) is 0.09 for the former sample compared to 
0.13 for the latter one.  If one takes into account only the parent halos,
then the scatters are 0.09 and 0.12, respectively. Therefore the difference
in the scatter of \cnfw\ between the VOID and CLUSTER halo samples is
mainly due to a (global) environmental effect. The same applies for the scatters
of the other concentration parameters and $\rho_{-2}$.

The concentrations and $\rho_{-2}$ of the FIELD (parent) halos 
tend to decrease with mass, but the trends are shallower than for the 
CLUSTER halo sample. 
We notice that the trends of $\log$\cnfw\ and $\log$\cq\ with $\log$\mv\ for 
our FIELD sample are shallower than those reported for isolated halos in previous 
works, where $\sigma_8=1$ was used \citep{AR99,Bullock2001b}, at least in the 
range of masses explored here. In Figs. 1 and 2 we show a toy-model prediction 
for \cnfw\ and $\rho_{-2}$ (long-dashed lines) following \citet{ENS2001} 
for the same cosmological parameters than in our simulations and $\sigma_8=0.9$. 
The toy model for \cnfw\ was normalized to a set of isolated halos in a wide 
range of masses (re)simulated with very high resolution \citep{ENS2001}.  
The toy model roughly agrees with our FIELD halo results.

Interestingly, studies aiming at infering the halo \cnfw\ concentration
from galaxy observables show that the slope of the $\cnfw-\mv$
relation is significantly steeper for halos of elliptical 
galaxies \citep[\eg][]{Sato00}, which are localized typically deep inside
the clusters,  than for halos of spiral galaxies \citep{Jimenez03},
which are localized in less dense environments. 

Our main interest here is to compare the properties of halos in low and 
high-density environments. From Figs. 1 and 2, one concludes that the 
concentrations and $\rho_{-2}$ of the VOID sample are lower 
on average than those of the CLUSTER sample, although the scatter in both 
samples is large. In fact this conclusion is based mostly on extrapolations 
of trends because the VOID and CLUSTER samples almost do not overlap in 
mass. At $\sim 10^{11}$\msunh, the concentrations \cnfw, \cq~ and \cd\ are on 
average 40\% smaller for halos in the VOID sample than for halos in the 
CLUSTER sample, and the inner density $\rho_{-2}$ is on average two times
lower for the former than for the latter. For larger masses, all these 
differences reduce. There are also differences in the scatters: they are 
larger for the CLUSTER sample than for the VOID and FIELD ones (see Table 3).

The halos from the FIELD sample also tend to have on average lower concentrations
and central density than those halos from the CLUSTER sample, although the
differences are smaller than those between VOID and CLUSTER halos, in 
particular for $\rho_{-2}$. 
If we assume that the prediction for \cnfw\ given by the \citet{ENS2001}
toy-model is a good fit to our FIELD sample data, then its extrapolation 
to lower masses shows that the difference in \cnfw\ between FIELD and CLUSTER halos 
remains roughly the same (Fig. 1). For $\rho_{-2}$, the difference seems to 
increase at lower masses (Fig. 2).

We have also compared only the parent halos from the CLUSTER sample with those
from the VOID and FIELD samples. The differences in concentrations
and $\rho_{-2}$ become less pronounced than in the case when the subhalos
were included. The slopes of the \cnfw, \cq, \cd\ and $\rho_{-2}$ vs mass
relations for the CLUSTER parent halos reduce to -0.07, -0.06, 
-0.07 and -0.16, respectively (to be compared with those given in Table 3),
becoming closer to those of the FIELD sample, but still steeper. As expected,
\cd\ changes more than \cnfw\ and \cq\ because the definition of the latter 
concentrations imply that the halos should be non-truncated, and the non-truncated
halos are mostly parent ones, so that the fraction of CLUSTER subhalos in Fig. 1 
is small in any case. 
We have also compared the concentrations and $\rho_{-2}$ of the very few 
subhalos from the VOID sample with the extrapolations of these parameters
to the corresponding masses from the CLUSTER subhalo sample. All the 
VOID subhalo concentrations and $\rho_{-2}$ lie below the corresponding
extrapolations from the CLUSTER subhalos.    

Our results show that halos in dense environments are on average more 
concentrated, with higher central densities and with larger scatters in these 
parameters than halos in low-density environments. We find that these
differences are owing to both nature and nurture reasons. Related to the former 
is mainly the fact that {\it halos in dense regions typically assemble most of their 
masses earlier than halos in low dense regions}, through violent MAHs.
\citet{ST04} indeed have shown that halos in dense regions form at earlier
times than do halos of the same mass in less dense regions. 
Related to ``nurture'' is the effect of steepening of the outer (sub)halo 
profile due to tidal stripping by the parent halo \citep{Ghigna98,OH99, AR99, 
Klypin99}. 
This effect tends to 
make the subhalos more concentrated and it is subject to a local condition
rather than to the global environment. Subhalos are indeed systematically more
concentrated than parent halos, however, both increase also their concentrations
as the environment density increases \citep[see also][]{Bullock2001b}. 
The effect of the formation epoch, related to the global environment, affects 
likely more the central halo density than concentrations, while the 
inverse is expected for the local evolutionary effects. As will be seen below, 
the differences with environment from $z=1$ to $z=0$ are more pronounced for 
concentrations than for $\rho_{-2}$.

 \begin{deluxetable}{lcccc}
\tablecolumns{4}
\tablewidth{0pc}
\tablecaption{Linear regression parameters (concentrations and $\rho_{-2}$)}
\tablehead{\colhead{Sample} & \colhead{lg$\cnfw=a+b$lg\mv} & \colhead{lg$\cq=c+d$lg\mv} & \colhead{lg$\cd=e+f$lg\mh} & \colhead{lg$\rho_{-2}=g+h$lg\mh} \\
   &  $a$\ \ \ \ \ \ $b$\ \ \ \ \ \ $\sigma$ &  $c$\ \ \ \ \ \ $d$\ \ \ \ \ \ $\sigma$ &   $e$\ \ \ \ \ \ $f$\ \ \ \ \ \ $\sigma$  &  $g$\ \ \ \ \ \ $h$\ \ \ \ \ \ $\sigma$ }
\startdata
\ \ \ $z=0$  \\
CLUSTER & 2.23\ \ -0.09\ \ 0.13 & 2.10\ \ -0.09\ \ 0.11 & \ 0.77\ \ -0.11\ \ 0.11 & 19.01\ \ -0.29\ \ 0.33 \\
VOID    & 1.10\ \ \ 0.00\ \ 0.09 & 0.96\ \ -0.00\ \ 0.07 & -0.38\ \ -0.01\ \ 0.07 &15.04\ \ \ 0.04\ \ 0.29 \\   
FIELD   & 1.51\ \ -0.04\ \ 0.11 & 1.53\ \ -0.05\ \ 0.08 & \ 0.00\ \ -0.05\ \ 0.10 & 15.94\ \ -0.05\ \ 0.32 \\
\ \ \ $z=1$  \\
CLUSTER & 1.40\ \ -0.05\ \ 0.15 & 1.36\ \ -0.05\ \ 0.09 &  -0.23\ \ -0.04\ \ 0.18 & 17.47\ \ -0.15\ \ 0.40 \\
VOID    & 1.15\ \ -0.02\ \ 0.10 & 1.22\ \ -0.04\ \ 0.06 &  -0.34\ \ -0.02\ \ 0.13 & 17.30\ \ -0.14\ \ 0.40 \\   
FIELD   & 0.67\ \ \ 0.00\ \ 0.15 & 0.60\ \ 0.01\ \ 0.09 & \ 0.15\ \ - 0.04\ \ 0.19 & 17.17\ \ -0.14\ \ 0.38 \\
\enddata
\end{deluxetable}

\subsubsection{Results at z=1}

We analyze the CLUSTER, VOID and FIELD samples at $z=1$ 
in the same statistical sense as done 
at $z=0$. Similar to Figs. 1 and 2, we show in
Figs. 3 and 4 the concentrations and $\rho_{-2}$ versus 
mass but now at $z = 1$, and in Table 3 the 
parameters of the corresponding 
linear regressions are given.  The first impresion is that the 
differences in the concentrations and central density seen at $z=0$
among the three halo samples are hardly present at $z = 1$.
Our results thus show that at this epoch the \cnfw, \cq, \cd\ and 
$\rho_{-2}$ versus mass relations do not depend on environment. 


\begin{figure}[htb!]
\vspace{11.5cm}
\includegraphics{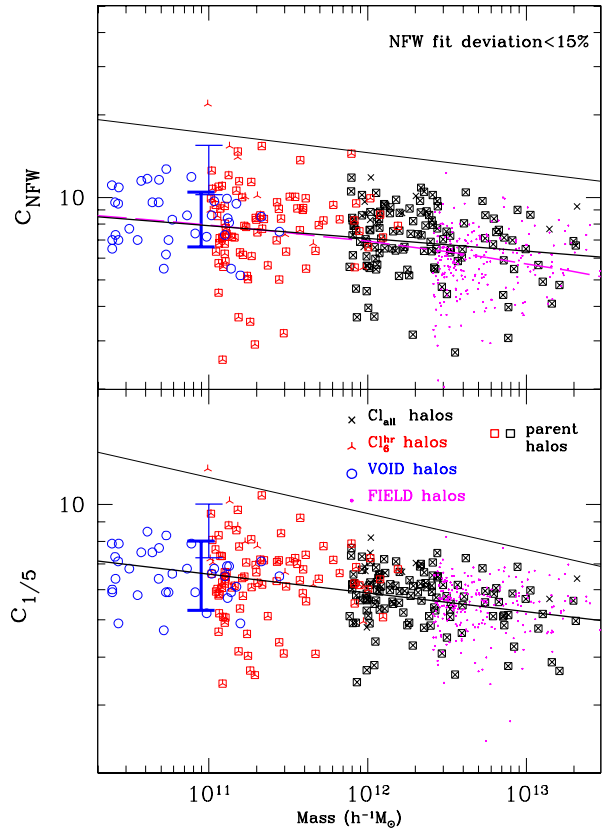}
\caption{Same as in Fig. 1 but for a snapshot at $z=1$. Thick lines and
error bar correspond to the current samples ($z=1$), while  thin line
and error bar are the same ones as shown in Fig. 1 ($z=0$). The bars
were slightly shifted to avoid overlapping.}
\end{figure}


\begin{figure}[htb!]
\vspace{11.5cm}
\includegraphics{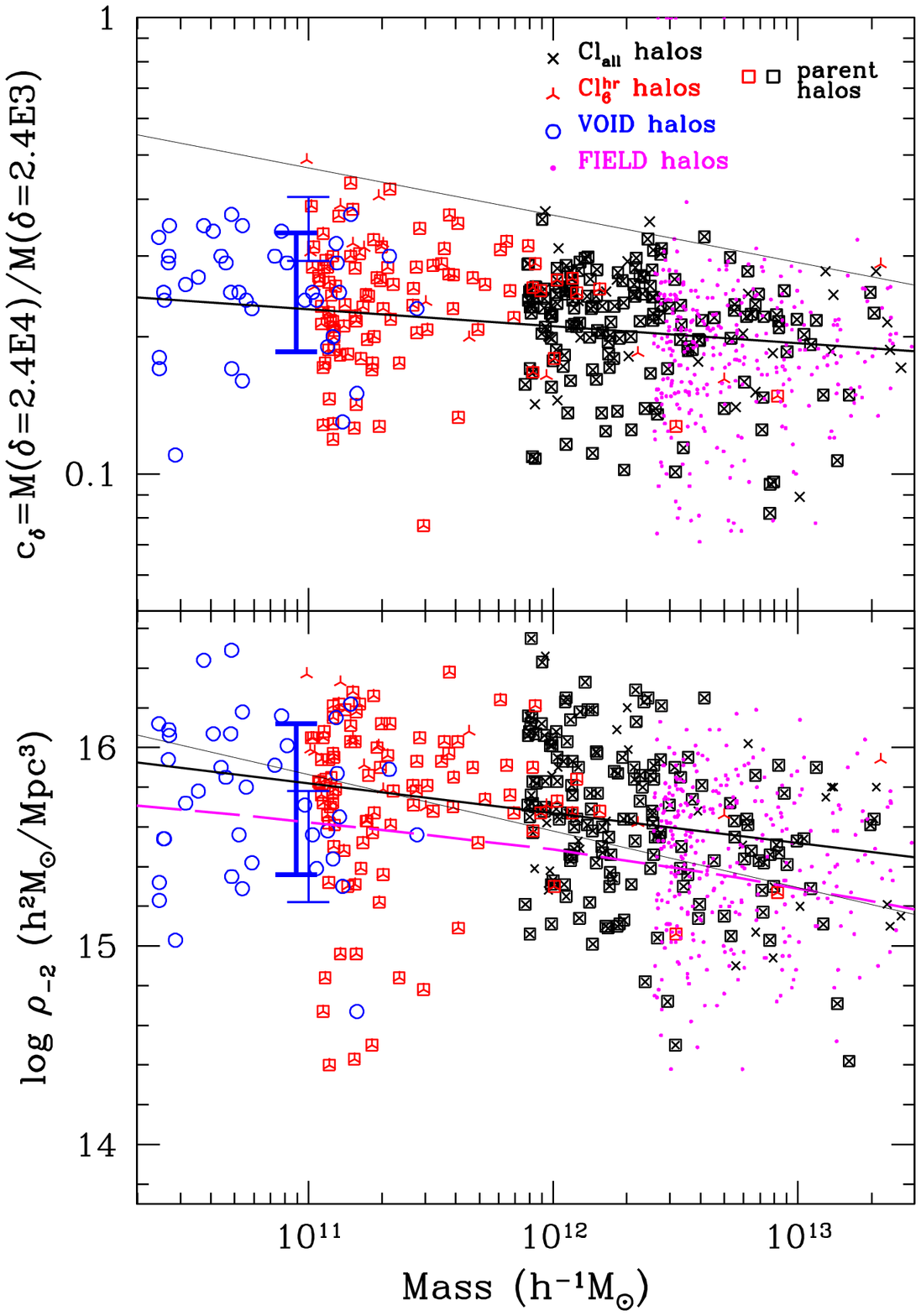}
\caption{Same as in Fig. 2 but for a snapshot at $z=1$. According
to the spherical collapse model, the overdensity is 1.65 times
smaller at $z=1$ than at $z=0$ for our cosmology; this was taken
into account in \cd. Thick lines and
error bar correspond to the current samples ($z=1$), while  thin line
and error bar are the same ones as shown in Fig. 2 ($z=0$).}
\end{figure}


The central concentration $\rho_{-2}$ for the CLUSTER halos is on average 
only a little larger at $z=1$ than at $z=0$. Halos in dense environments
assemble their present-day mass early in such a way that their central
densities are stablished likely before $z=1$. 
For the VOID halos, the mean of $\rho_{-2}$ at $z=1$ is larger than 
the one at $z=0$ by $\sim 70\%$. Halos in low-density environments
assemble slowly, incorporating most of their mass lately. 

The concentration parameters change more with $z$ than the central
density. As previously reported \citep[\eg][]{Bullock2001b, 
ENS2001, Navarro04}, the \cnfw\ concentration parameter 
is lower at higher redshifts for a given mass.  In the upper panel 
of Fig. 3, the dashed curve shows the $\cnfw-\mv$ dependence at 
$z=1$ given by the toy model of \citet{ENS2001}.
According to this model $\cnfw\propto (1+z)^{-1}$ approximately \citep[see 
also][]{Bullock2001b}.  In Figs. 3 and 4
are also shown (thin solid lines) the linear regression for the CLUSTER 
sample and the 1$\sigma$ scatter of the VOID sample plotted in Figs. 1 and 2 
($z=0$).  The concentration \cnfw\ is on average $\sim 1.5$ and 2.1 times
lower at $z=1$ than at $z=0$ for the VOID and CLUSTER samples, respectively. 
For the latter sample, the comparison is made at $10^{12}\msunh$. For the 
FIELD sample, the differences of \cnfw\ between $z=0$ and $z=1$ are
on average slightly less than a factor of two, at least in the range
of masses explored here. Thus, the evolution of \cnfw\ seems to be
slightly different in the different environments. A similar behaviour
is seen for \cq\ and \cd. The scatters in all the measured parameters 
are larger at $z=1$ than at $z=0$, reflecting likely the fact that at $z=1$ 
the halos are in general less relaxed than at $z=0$.

The increase of concentration with time is more pronounced in 
the cluster environment. We analyze the behaviour of concentration for 
halos inside the present-day cluster virial radius (subhalos) in simulation 
\clh. Because for this simulation we have the halo MAHs, concentrations 
for a given halo can be measured at $z=0$ and $z=1$. 
We find that the ratios of concentrations measured at $z=0$ 
and $z=1$ tend to increase as the $\mh(z=0)/\mh(z=1)$ ratio decreases
(Fig. 5); i.e., halos that have suffered more mass loss due to tidal 
stripping, end preferentially with higher concentrations.
Therefore, as mentioned above, tidal stripping seem
to be an efficient mechanism for increasing the average concentration
of halos. Moreover, since the concentration is an increasing
function of the mass loss, then some steepening of the concentration-mass
relation in high-density environments is expected. This is what
we observe in our simulations, in particular for \cd\ and \cq\ (Figs. 3 and 4). 
For \cnfw\ this effect is not seen because only halos that have a good
fit to the NFW profile are selected, and these are typically
well relaxed halos that have not felt strong tidal stripping. 

\begin{figure}[htb!]
\plotone{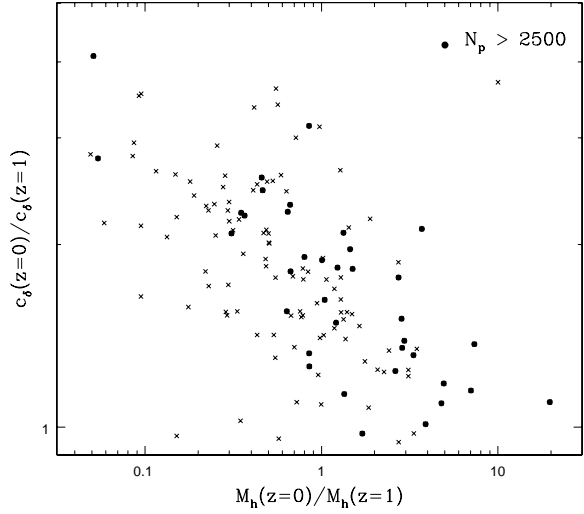}
\caption{Ratio of the concentration parameter \cd\ measured at
$z=0$ and $z=1$ vs the ratio of the halo masses at these epochs
for individual halos inside the present-day virial radius of cluster 
\clh. Halos with more than 500 particles at $z=0$ are shown. Those
with more than 2500 particles are marked with solid circles. 
Halos that have lost more mass, i.e. that underwent more 
tidal stripping, have systematically increased more their concentrations
than halos less affected by tidal stripping.}
\end{figure}

\subsection{Halo shape distribution}

Halo axis ratios as well as the directions of the principal axes 
are determined by iteratively\footnote{The first iteration consider
all particles inside a sphere of radius equal to the halo radius. 
For the second iteration, we reorient the system so as the $x$ axis
lies on the major axis of the ellipsoid found in the first iteration. 
We now use all particles inside the ellipsoid. 
The iteration proceed untill convergence is obtained.} 
diagonalizing the tensor
\begin{equation}
I_{ij} = \sum x_i x_j/ a^2, \\\\\ a \equiv \sqrt{x^2 + \frac{y^2}{q^2}
+ \frac{z^2}{s^2}}.  
\end{equation}
The sum is over all particles within the halo radius \rh\ (virial or 
truncation radius, whichever is smaller), $x_i$
($i=1,$ 2, and 3) are the particle coordinates with respect to 
the halo center of mass, $s$ is the short-to-long axis ratio, and $q$
the intermediate-to-long axis ratio. We use here halos with 
more than 500 particles. 

To obtain an estimate of the error in the determination of the axial ratios 
due to low particle number, we analyze the well-resolved cluster halos from 
our simulations. For each one of these halos, which contain hundreds of 
thousands of particles, we randomly extract only 
500 particles and measure $q$ and $s$ in 100 realizations. 
We found the the 1$\sigma$ errors in $q$ and $s$ 
amount to about 10\% and that they do not depend on the mass 
of the cluster-size halo. These errors are an upper limit because our 
galaxy-size halos have more than 500 particles.

\begin{figure}[htb!]
\plotone{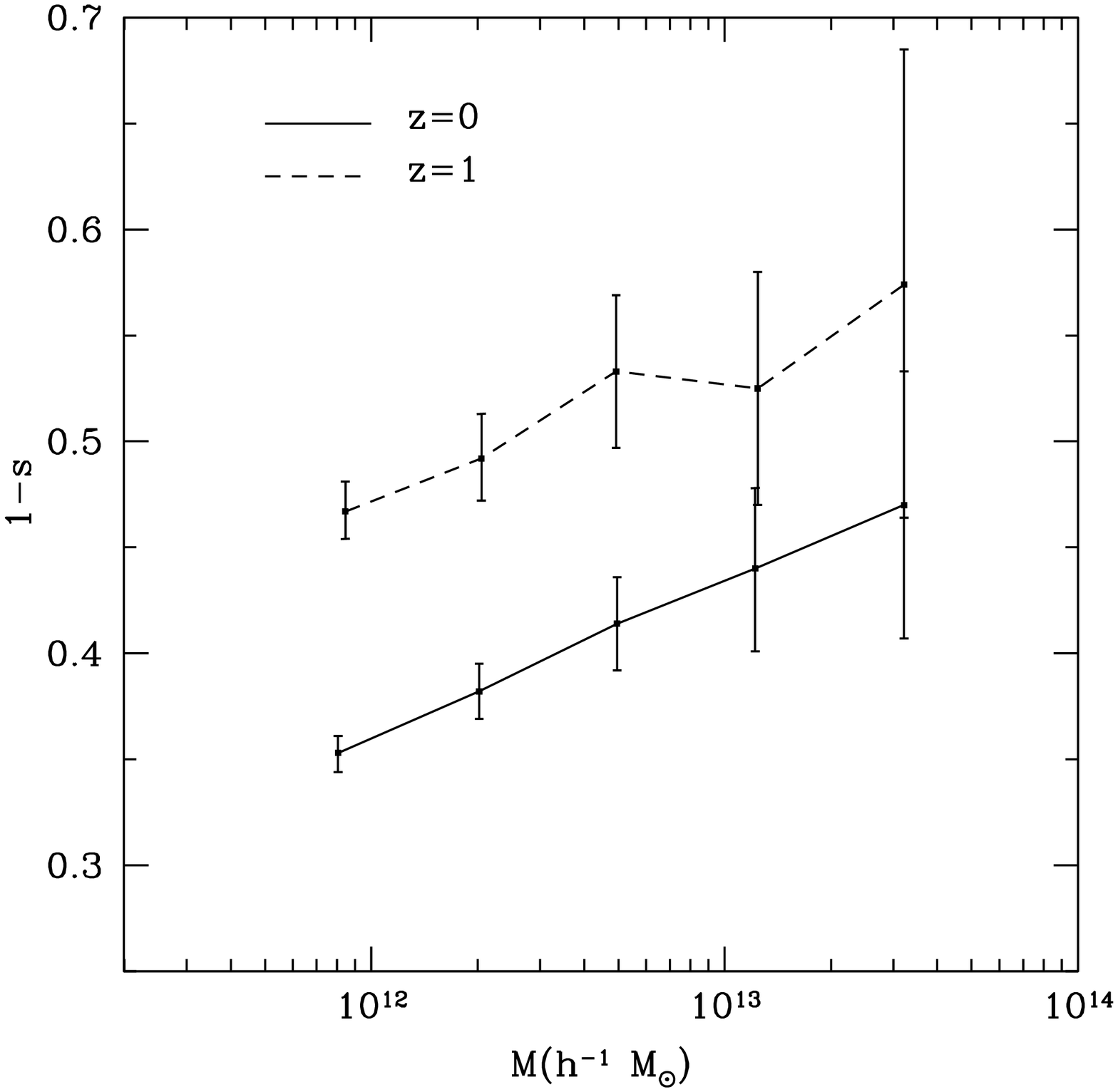}
\caption{The average ellipticity, $\epsilon = 1-s$, as a function of 
halo mass at two redshifts $z =0$ and 1 for halos in the FIELD
sample. Error bars reflect the Poisson 
uncertainty associated with the number of halos within the bin and not
the scatter in the relation.}
\end{figure}


\begin{figure}[htb!]
\plotone{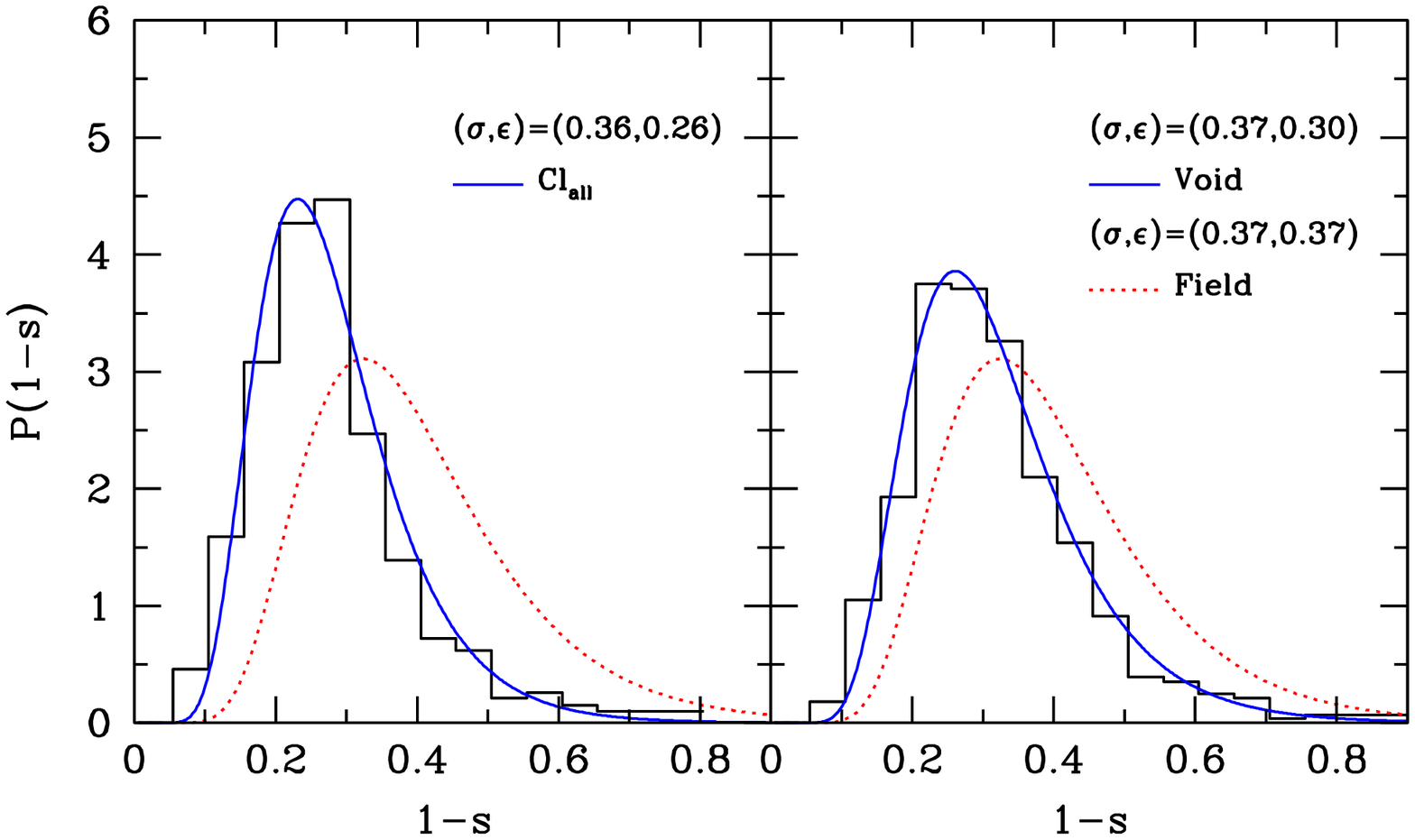}
\caption{Ellipticity distribution (histograms) 
for halos with more than 500 particles drawn from a cluster environment
(left panel) and from a void environment (right panel). Curves
on each panel are lognormal best-fits to the data. The lognormal
fit of the $\epsilon$ distribution for the FIELD sample of halos 
is shown in both panels for comparison purposes. The values of the 
parameters of the fits $(\sigma_{\epsilon},\epsilon_0)$ are also 
shown in panels.}
\end{figure}


Figure 6 shows the ellipticity as a function of halo mass at two redshifts
for the FIELD halo sample. For a given epoch, low-mass halos 
are preferentially more spherical than high-mass halos while,
for a given mass, halos at $z=1$ are on average 
more flattened than halos at $z=0$. These results
agree with those of \citet{Bull01} and \citet{HBB05} and seem
to suggest that present-day halos, having had more time to relax
and attain a more isotropic orbit distribution, 
as compared with high-redshift ones, tend to be more spherical.
Moreover, since small halos form earlier on average than big ones, 
they also have had more time to relax and thus become more spherical.

In Fig. 7 we show the distributions of the halo ellipticity 
$\epsilon_1 \equiv 1-s$ for our different samples. The ellipticity 
distribution of sample \cll\ is shown in left panel while the one
for sample VOID is shown in right panel. They are both well fit by a 
lognormal distribution,
\begin{equation}
P(\epsilon) = \frac{1}{\epsilon\sqrt{2\pi} \sigma} \exp 
\left( - \frac{\ln^2(\epsilon/\epsilon_0)}{2 \sigma^2} \right),
\end{equation}
with best fit values ($\sigma$,$\epsilon_0$) given on the 
corresponding panels. Curves on
panels are the lognormal best fits to the
data. The lognormal fit of sample FIELD is shown as a dotted line in both 
panels for comparison purposes. 

A Kolmogorov-Smirnov (KS) test shows that the probability
that \cll\ and FIELD halo samples are drawn from the same 
parent distribution
is very small, $3.0 \times 10^{-6}$. Thus, according to Fig. 7, halos in
clusters and their outskirts are on average more spherical 
than halos in voids or in the field. This conclusion may result at first
at odd because, being halos in clusters more massive on average 
than their counterparts in voids, they are expected to be more
flattened (see Fig. 6). However, the {\it environmental} effect
related to the halo formation epoch seems to dominate over the
weak dependence of ellipticity on mass. Halos in a high-density 
environment assemble earlier than in a low-density one, and
as follows from Fig. 6, halos tend to be more spherical as they
assemble earlier.  On the other hand, when the comparison is made 
between halos from the VOID and FIELD samples, the  mass effect 
seems to dominate over the environmental one, in such a way
that the former are on average less flattened than the latter.

\section{Spin parameter and angular momentum alignment}

\subsection{Spin parameter distribution}

We compute the halo total angular momentum as
\begin{equation}
\mbox{\boldmath $J$} = \sum_{i=1}^{n} m_i \mbox{\boldmath $r_i$} 
\times \mbox{\boldmath $v_i$},
\end{equation}
where \mbox{\boldmath $r_i$} and \mbox{\boldmath $v_i$} are the 
position and velocity of the {\it i}th particle with respect to the halo
center of mass. We follow \citet{Bullock2001a} and define
a modified spin parameter  $\lambda'$ to characterize the 
global angular momentum of a halo
\begin{equation} 
\lambda' \equiv \frac{J_h}{\sqrt{2} M_h V_h R_h},
\end{equation}
where $J_h$ is the angular momentum inside the halo 
radius \rh\ (see \S 3.1), and $V_h$ is the circular 
velocity at \rh. 
Hereafter, in this section, we drop the prime in $\lambda'$.


\begin{figure}[htb!]
\plotone{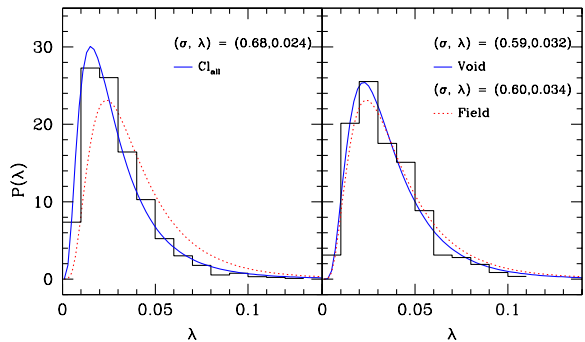}
\caption{Spin parameter distribution (histograms) for halos 
with more than 500 particles drawn from a cluster environment
(\cll\ sample, left panel) and from a void environment (VOID sample, 
right panel). Curves on each panel are lognormal best fits to the data. 
The lognormal fit of the $\lambda$ distribution for halos in the FIELD sample 
is shown in both panels for comparison purposes. The values of the 
parameters of the fits ($\lambda_0,\sigma_{\lambda}$) are also shown in panels.}
\end{figure}


Figure 8 shows the spin parameter distributions for our different galaxy-size 
halo samples. In this case, halos with more than 500 particles were used. 
The halos that are subhalos are cut at the truncation radius. 
The spin distributions of the \cll\ and VOID samples
are shown in left and right panels, respectively. 
Curves on panels are the corresponding lognormal best fits 
to the data. The lognormal fit of sample FIELD 
is shown as a dotted line in both 
panels for comparison purposes. The ($\sigma$,$\lambda_0$) parameters
of the lognormal distributions are given inside the panels. The median 
values of $\lambda$ for the different samples are presented
in Table 2. The conclusion from Fig. 8 and Table 2 is clear: 
halos from the \cll\ sample have on average lower spin parameters 
and a wider distribution than halos from the VOID and FIELD samples. 
The $\lambda_0$ and $\sigma$ parameters of the former sample are 
30-40\% and $\sim 15\%$ times smaller than for the latter samples,
respectively.

The difference between the $\lambda$ distributions of halos in the 
\cll\ and VOID samples is confirmed by a KS test: the probability
that both samples belong to the same parent distribution is
$1.17\times 10^{-5}$. Our result agrees with that obtained by 
\citet{reed04}. They also found that halos in high-density 
environments have smaller spins than halos in the field. Are these
differences due to global environment effects or to the fact that CLUSTER
subsamples have a higher fraction of subhalos? To address this question
we compare the $\lambda$ distributions of the parent halos and subhalos
of the CLUSTER subsamples. By applying a KS test we find only slight
differences in these distributions: the probabilities that both 
samples belong to the same parent distribution are 0.89 for the \clh\ 
sample and 0.45 for the \cll\ sample. On the other hand, the $\lambda$ 
medians (or  $\lambda_0$) of the halo and subhalo samples in both cases 
agree within 5\%. 
We notice that the FIELD sample is by construction composed of only parent 
halos, and the VOID halos are esentially parent ones (more than 90\%).

How much of the difference we have found can be attributed to 
cluster-to-cluster scatter? Unfortunately, we only have one
cluster, \clh, with hundreds of halos, each one with more than 500 
particles inside its virial or tidal radius. The other clusters are less
resolved and they thus have less halos than \clh\ by far.
We avoid the halo {\it low-number} problem by combining these
clusters to form only one, which in principle is
independent from \clh. We measure $P(\lambda)$ for the 
independent \clh\ halo sample
and test whether this distribution is similar to the one 
obtained for the composite \cll\ sample.
We found differences in the values of the fitting parameters 
($\sigma$,$\lambda_0$) of less than 5\%. Slightly higher 
differences are measured when the comparison is made
between the $P(\lambda)$ distribution from the same sample 
\cll\ but at two close snapshots.

Our results show that present-day halos in dense environments 
have a spin parameter distribution shifted to lower values with respect
to those of halos in the field or in the voids. 
Does $\lambda$ decrease due to the lost of high angular 
momentum material by halos that suffer strong tidal stripping in
dense regions? A way of testing this hypothesis is 
by simply measuring $P(\lambda)$ in the cluster environment 
at much earlier time, when most halos have yet to experiment strong 
tidal stripping. We find a median value $\lambda_{med} = 0.036$ at 
$z = 1$ for the CLUSTER sample, which is comparable to 
$\lambda_{med} = 0.033$ for the FIELD sample at present time. This 
result is consistent with the hypothesis above but does not 
actually prove it. On the other hand, we showed above that subhalos and
parent halos have, according to a KS test, similar $P(\lambda)$
distributions. These results appear to contradict each other. Notice,
however, that the histories of parent halos in high-density regions can 
be very different from their counterparts in low-density environments because
the former undergo a ``tumultuous life''. They may have suffered, for instance, 
tidal stripping in the past, from close encounters with major substructures, 
or their mass accretion could have been stopped \citep{Kravtsov04b}. 
We also looked for a cluster-centric radial 
$\lambda$ dependence in \cll\ and \clh\ samples. Unlike \citet{reed04}, we did 
not find any systematical decreasing of $\lambda_{med}$ as the cluster-centric 
radius is smaller. Notice that \citet{reed04} analyzed subhalos with more than 
144 particles, while our subhalos have more than 500 particles; many subhalos 
are below this limit, but the measure of $\lambda$ for them is not reliable.
 
For the VOID and FIELD samples, $\lambda_{med} = 0.034$ and 0.036 at $z=1$, 
respectively. Therefore, $\lambda$ for these halo samples also decreases on
average from $z=1$ to $z=0$, but very little. In this case, the small decreasing 
of $\lambda$ with time could be explained by the accretion mode proposed by 
\citet{pmdf04}. In summary, while halos in low-dense regions seem to keep
constant or slightly decrease their $\lambda$ values from $z=1$ to $z=0$, 
halos from the cluster-like regions tend to decrease significantly their 
$\lambda$ values. We interpret the latter as a consequence of two effects: 
(i) the earlier halo assembly epochs typical of higher density regions, and (ii) 
the subhalo tidal stripping and the the tumultuous mass assembly 
history of halos in clusters and their surroundings. For a more detailed 
discussion of these 
effects it is necessary to follow the individual mass and angular momentum assembly 
histories of the halos. In a subsequent paper we will construct and analyze these 
individual histories for each one of our samples.

\subsection{Internal angular momentum alignment}

In previous subsection we explored the distribution of the 
magnitude of the angular momentum as a function of environment. 
Now we will explore possible differences in the distribution of the angular
momentum internal alignment. We measure the typical alignment
by the angle \teta\ between the mean angular momentum of the
particles in the inner sphere of half-mass radius and the mean 
angular momentum of the particles in the outer half-mass spherical 
shell \citep{Bullock2001a}. As discussed by \citet{Bullock2001a} large
errors are involved in the determination of the direction of the
angular momentum vectors and thus in \costha. The error in \costha\ is 
estimated in two manners. In the first one, we assign errors
to each component of the angular momentum, in both inner and outer
half-mass regions, using a Monte Carlo procedure similar to the one 
built to estimate the error in the axial ratios (\S 3.2), and 
perform a standard propagation of errors. In the second
one, the Monte Carlo procedure is applied directly on \costha.
The first method gives larger errors. This is so likely because the 
internal and external angular momentum are not entirely independent 
quantities. We found that the error depends not only on the number of 
particles, $N_p$, and $\lambda$ but also on the intrinsic
alignment: a more aligned halo have a smaller error. For $N_p$ as low
as $\sim 500$, the $1\sigma$ error in \costha\ can reach 50\%
of the measurement. {\it Bearing this in mind}, we now proceed to present 
our results (still uncertain) and their possible intepretations. 

The distributions of $\cos\theta_{1/2}$ for our different environments 
at $z = 0$ are shown in Fig. 9. As it was the case for $P(\lambda)$, we use 
halos with more than 500 particles. 
Figure 9 shows that the \cll\ halos 
are on average less aligned than the halos from 
the FIELD and VOID samples. The median values of $\cos\theta_{1/2}$
for the \cll\, FIELD, and VOID samples are 0.51, 0.80, and
0.83, respectively. In order to find a possible explanation for these
differences with environment, we have also measured the alignment 
distributions at a much earlier time, $z = 1$ (Fig. 10). Figures 9 and 10 
show that the population of VOID and FIELD halos at $z=0$ is more aligned 
on average than the one at $z=1$. A more quantitative estimate of this 
difference is confirmed by the KS test: for the FIELD sample, for example, 
the probability that the populations at $z = 0$ and $z = 1$ belong to the 
same parent distribution is $4.5 \times 10^{-5}$, with $\cos\theta_{\rm 1/2,med} = 0.80$ 
and 0.75, respectively. Regarding the \cll\ sample, there is no any 
significant difference in the alignment distributions of halos at $z=0$ 
and $z=1$.


\begin{figure}[htb!]
\plotone{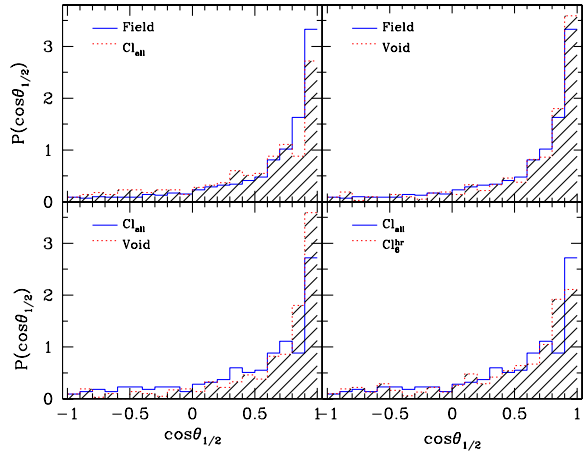}
\caption{Angular momentum alignment distributions for our 
four samples of halos at $z=0$. Two different distributions 
are shown in each 
panel for comparison purposes. Distributions shown as 
dotted histograms are also shaded for more clarity. 
Halos in voids and those in the field present similar 
distributions but they both are on average more aligned 
than their counterpart in clusters.}
\end{figure}


We remark that the results presented above need to be confirmed 
by future analysis with better resolved halos. If confirmed, we
find the following explanation for the evolutionary 
effect seen in the internal angular momentum alignment of FIELD and VOID 
sample halos. On one hand, most of the angular 
momentum of the outer sphere, which is where essentially
most of the halo angular momentum resides, was acquired during 
the linear grow. On the other hand, it is expected that most of the angular 
momentum of the inner sphere comes from the violent, initial
merger-growth phase \citep{maya02, pmdf04}. 
The material accreted by the halo at later
times brings angular momentum oriented in 
the direction of the angular momentum of the outer shell. 
As part of this material is incorporated into the inner
sphere the alignment increases. This mechanism is not
expected to apply to halos in clusters or their outskirts because 
they typically do not incorporate material as soon as they fall 
into the cluster halo; in some cases the halos in clusters
even loose mass due to tidal stripping.  

Regarding the question of why present-day halos in clusters 
are less aligned than those in the field or in voids, a
comparison between \cll\ and FIELD halos at $z = 1$ shows that 
they are equally well aligned. This is probably so because at 
$z = 1$ these environments are still not too disimilar.
However, unlike halos in the field and voids, halos in 
clusters evolve under the influence of
strong tidal fields. Halos in sample CLUSTER
have a mass assembly history very different from their
counterparts in samples FIELD or VOID. In the cluster
environment, a significant fraction of halos present at 
$z = 1$ do not survive until $z = 0$, others end up with their
masses significantly reduced, while some others, those that at 
present day are at cluster outskirts, may grow as what is typical 
of a halo in field and void environments.
Figures 9 and 10 show that halos in sample \clh\ at $z = 0$ are
more disaligned than halos at $z = 1$. For the subset of halos
of sample \clh\ that are within \rv\ at $z = 0$ (subhalos), 
which were also identified at $z = 1$,
we find the same trend: the halos at $z = 1$ are on average more
aligned than the halos at $z = 0$. A halo by halo 
comparison, however, shows that the trend is not systematic; i.e., 
there are halos that exhibit a higher alignment at $z = 0$. 
In summary, it seems that in an environment
in which halos stop growing, an increase in alignment
is not expected, but even more, it seems that a cluster-like
environment acts on the halo angular momentum internal alignment
in a non-trivial way: the change in alignment varies halo by
halo. 


\begin{figure}[htb!]
\plotone{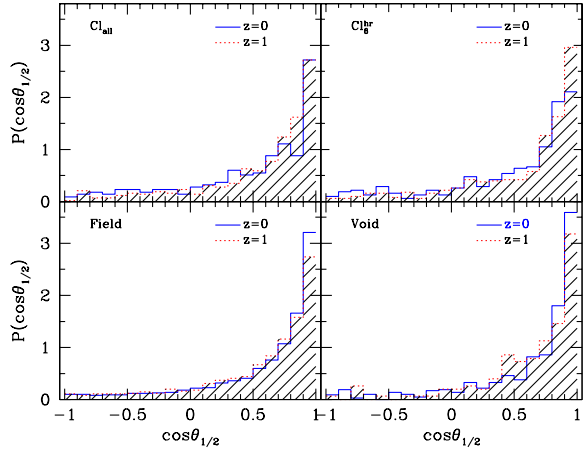}
\caption{Angular momentum alignment distributions for our 
four samples of halos at $z=0$ and $z=1$. Each panel 
correspond to a halo sample. Distributions shown as 
dotted histograms are also shaded for more clarity. 
Halos in voids and those in the field are on
average less aligned at $z = 1$ than at $z = 0$.}
\end{figure}


\subsection{Alignment between the halo shape and angular momentum}

Halo axis ratios as well as the directions of the principal axes 
are determined as explained in \S 3.2. We denote with $\theta$
the angle between the angular momentum axis within \rh\ and the 
direction of the minor principal axis. In Fig. 11 we show the
distributions of $\theta$ at $z = 0$ for our different halo
samples. We use here also halos with more than 500 particles, 
although the iterative procedure 
sometimes need more than that to converge. The number of halos used 
to draw the histograms in Fig. 11 can thus be lower than those
shown in Table 2 (see \S 2). We repeat here the procedure we follow
to evaluate the $1\sigma$ error in axial ratios and \costha\ 
to obtain the corresponding $1\sigma$ error in \costh. Errors in 
\costh\ are smaller than those ones in \costha. They amount to about
20\% for halos with 500 particles. Taking 
into account that this particle number is a lower limit, we expect
errors in \costh\ to be on average less than 20\%. 
On the other hand, errors in \costh\ are
also dependent on the degree of alignment but they
seem to be less sensitive to it than the errors in \costha. 

The angular momentum axis of halos tends to align to the 
minor axis of the halo ellipsoids (in a plot like Fig. 11, a random 
oriented distribution would be uniform). This appears to be a generic 
prediction of the CDM cosmology \citep[see also][]{FGKM02,bs04}. Besides, 
we find some trend of decreasing the alignment angle $\theta$ from \cll\ 
to FIELD and VOID halo samples. This visual result is mildly supported by 
a KS test: the probability that the \cll\ and VOID samples belong to the 
same parent distribution is $2.4 \times 10^{-2}$. 

The relative orientations of the halo angular momentum direction and the 
halo principal axes could have important consequences on disk galaxy 
formation, the dynamics of the disks inside the halos as well as on
observational studies aimed to explore the distribution of satellite
galaxies with respect to the parent halo shape and orientation.


\begin{figure}[htb!]
\plotone{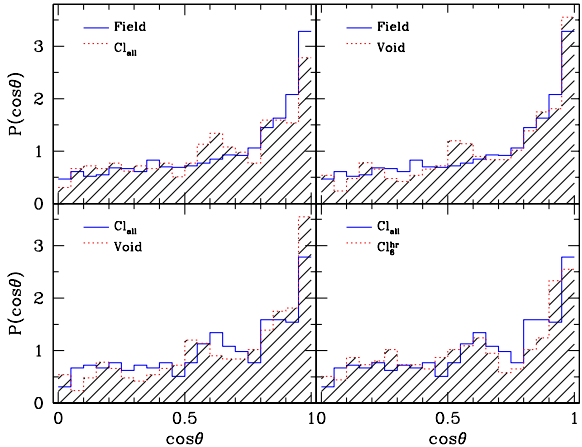}
\caption{Distributions of the alignment of the minor 
principal axis and the angular momentum within \rh\ for our 
four samples of halos at $z=0$. Two different distributions 
are shown in each 
panel for comparison purposes. Distributions shown as 
dotted histograms are also shaded for more clarity. 
Halos in voids and those in the field present similar 
distributions but they both are on average more aligned 
than their counterpart in clusters.}
\end{figure}


\section{Structural relations}

\subsection{Maximum circular velocity vs mass}

Figure 12(a) shows maximum circular velocity, \vm, vs. mass, \mh, for 
the halos from all the samples analyzed here. 
As usual, \vm\ is the maximum value of the smoothed radial profile 
$V_{\rm circ}(r)=\sqrt{GM(r)/r}$, computed under the assumption of spherical 
symmetry.  
The solid and dashed lines 
are the linear regressions for the CLUSTER and VOID samples, respectively. 
The Pearson correlation coefficients of the CLUSTER, VOID, and FIELD 
samples are 0.98, 0.97, and 0.96, respectively. The parameters
of the linear regressions in the logarithmic plane are given in column
(2) of Table 4. 

  
\begin{figure}[htb!]     
\vspace{11.5cm}
\includegraphics{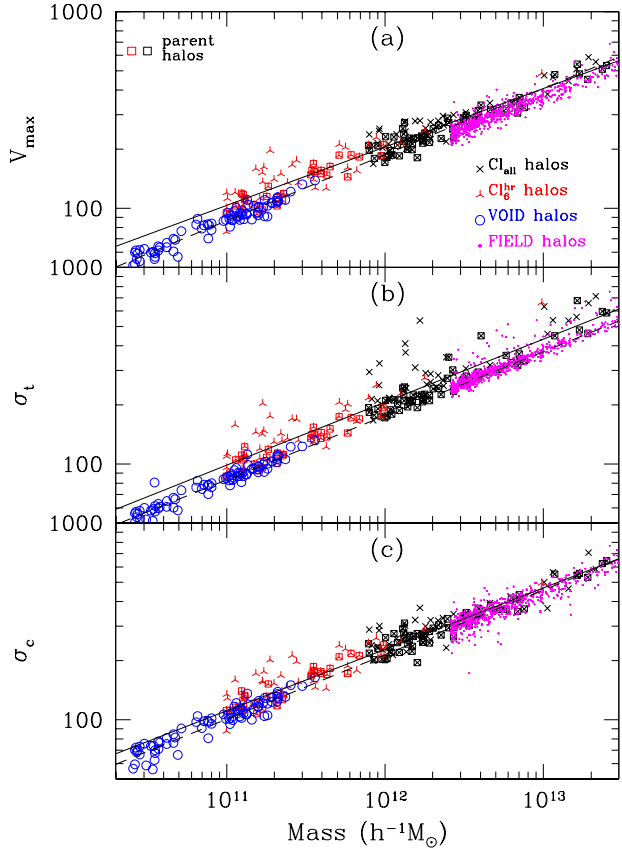}
\caption{Maximum circular velocity (a), and 3D velocity dispersions averaged
within \rh\ (b)and 0.1\rh\ (c) vs. halo mass for the \cll, \clh, VOID 
($V_{\rm all} + V_{50}^{hr}$), and FIELD samples. Solid and dashed lines
are linear regressions to the CLUSTER and VOID samples, respectively.}
\end{figure}


 \begin{deluxetable}{lccc}
\tablecolumns{4}
\tablewidth{0pc}
\tablecaption{Linear regression parameters (\vm\ and rms velocities)}
\tablehead{\colhead{Sample} & \colhead{lg$\vm=a+b$lg\mh} & \colhead{lg$\sigt=c+d$lg\mh} & \colhead{lg$\sigc=e+f$lg\mh} \\
   &  $a$\ \ \ \ \ \ $b$\ \ \ \ \ \ $\sigma$ &  $c$\ \ \ \ \ \ $d$\ \ \ \ \ \ $\sigma$ &   $e$\ \ \ \ \ \ $f$\ \ \ \ \ \ $\sigma$ }
\startdata
\ \ \ $z=0$  \\
CLUSTER & -1.25\ \ 0.297\ \ 0.05 & -1.52\ \ 0.320\ \ 0.08 & -1.39\ \ 0.312\ \ 0.06 \\
VOID    & -1.76\ \ 0.336\ \ 0.03 & -1.67\ \ 0.326\ \ 0.03 & -1.60\ \ 0.328\ \ 0.04 \\   
FIELD   & -1.76\ \ 0.334\ \ 0.03 & -1.71\ \ 0.327\ \ 0.04 & -1.54\ \ 0.322\ \ 0.04 \\
\ \ \ $z=1$  \\
CLUSTER & -1.55\ \ 0.324\ \ 0.04 & -1.74\ \ 0.343\ \ 0.07 & -1.43\ \ 0.319\ \ 0.06 \\
VOID    & -1.53\ \ 0.321\ \ 0.02 & -1.52\ \ 0.321\ \ 0.02 & -1.48\ \ 0.323\ \ 0.04 \\   
FIELD   & -1.77\ \ 0.340\ \ 0.03 & -1.82\ \ 0.370\ \ 0.05 & -1.32\ \ 0.311\ \ 0.05 \\
\enddata
\end{deluxetable}

Our results confirm the tightness of the \vm-\mh\ relation,
in this case, for different environments. The relation is steeper
for the VOID and FIELD samples than for the CLUSTER one. 
This is in agreement with the fact that the concentration-mass
dependences for the former are shallower (almost absent) than for
the latter (see below). In \citet{AR99} it was shown
how \vm\ does depend on \mh; for example, for the NFW model (their eq. 8)
$\vm\propto\mv^{1/3}g(\cnfw)$, where $g(\cnfw)$ is a function of 
\cnfw\ and \cnfw\ depends on \mv. In the limiting case
of no dependence of \cnfw\ on \mv, $\vm\propto\mv^{1/3}$, which
is close to our results for VOID and FIELD halos. The steeper the 
dependence of \cnfw\ on \mv, the shallower the \vm-\mv\ relation.

As seen in Fig. 12(a), for a given mass, halos in voids and in 
the field have smaller \vm\ than halos in clusters. At $10^{11}\msunh$
($3\ 10^{12}\msunh$), the circular velocity of VOID (FIELD) halos is 
typically $\sim 20\%$ ($\sim 12\%$) smaller than the corresponding to 
CLUSTER halos. On the other hand, the CLUSTER sample shows the largest
scatter in the \vm--\mh\ relation. Several authors who modeled disk
galaxy formation and evolution have shown that the Tully-Fisher relation 
(TFR) of disk galaxies is well explained mainly as an imprint of the CDM halo 
\vm--\mh\ relation \citep[\eg][]{MMW98, SN99, 
FA00, AF00, BJK01}. Thus, our results would imply that the TFR in different
environments could be slightly different (see also \S 6).  

In Fig. 13(a) we present the plot corresponding to Fig. 12(a) but at $z=1$. 
For comparison purposes, the same linear regresions shown in Fig. 
12(a) (solid and dashed thin lines) are plotted. The 
\vm--\mh\ relations in the different environments
are more similar at $z=1$ than at $z=0$. For a given mass, 
halos at $z=1$ have on average slightly larger \vm\ than at $z=0$,
specially for the VOID halos. This is because in a hierarchical
scenario \mh\ increases with time typically more than \vm\ does,
shifting the relation to the high mass (low velocity) side.
Nevertheless, the hierarchical mass growing can be stopped or even 
reversed in the cluster environment.   
The parameters of the linear regresions for the CLUSTER, 
VOID, and FIELD samples at $z=1$ are given in column (2) of Table 4.
The scatter of the \vm-\mh\ relation is smaller at $z=1$ than at $z=0$, 
specially for the CLUSTER sample.


\begin{figure}[htb!]
\vspace{11.5cm}
\includegraphics{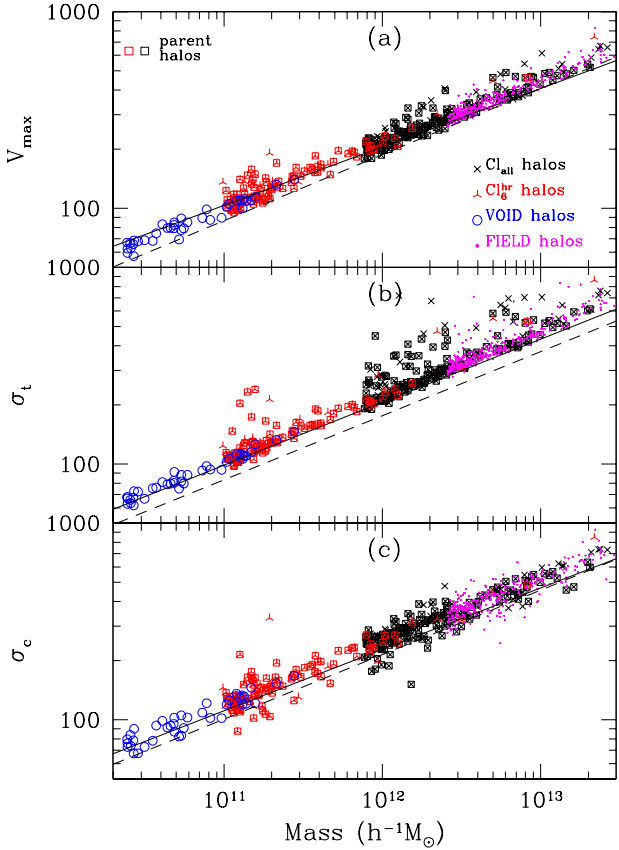}
\caption{Same as in Fig. 11 but for a snapshot at $z=1$. For comparative
purposes, the linear regressions of the CLUSTER and VOID samples at $z=0$
(Fig. 11) are shown (thin solid and dashed lines, respectively). }
\end{figure}


\subsection{Velocity dispersion vs mass}

Panels (b) and (c) of Figs. 12 and 13 show mass-weighted halo velocity 
dispersions vs mass for all samples at $z=0$ and $z=1$, respectively. 
The total and central velocity dispersions are defined as follows:
\begin{eqnarray}
\sigt^2=\frac{4\pi\int_0^{\rh}\sigma^2(r)\rho(r)r^2dr}{M(<\rh)} \cr 
\sigc^2=\frac{4\pi\int_0^{0.1\rh}\sigma^2(r)\rho(r)r^2dr}{M(<0.1\rh)},
\end{eqnarray}
where $\sigma(r)$ is the 3-dimensional velocity dispersion
at the spherical shell with radius $r$. The linear fitting parameters
for all the samples are given in third and fourth columns of Table 4.
The behaviour of the velocity dispersion-mass relations at $z=0$ 
for all the samples (environments) is similar to the one of
the maximum circular velocity-mass relation (see also Antonucci-Delogu
et al. 2001), reflecting this that most of the halos are close to their 
equilibrium state. In the plots
for \sigt\ vs \mh\ there are several outliers in the CLUSTER sample. These
are mostly truncated halos. They are not outliers in the \sigc\ vs \mh\
plot. Halos in voids and in the field have systematically smaller velocity
dispersions than halos in clusters, although the difference becomes
very small when the central velocity dispersion is used.

 \begin{deluxetable}{lcccccccccc}
\tablecolumns{4}
\tablewidth{0pc}
\tablecaption{Main properties of simulated disk galaxies}
\tablehead{\colhead{$M_d$} & \colhead{Env.\tablenotemark{a}} & \colhead{\cnfw} & \colhead{$\lambda$} & \colhead{$\lg\Sigma_0$} & \colhead{$f_g$} & \colhead{$B-V$} & \colhead{b/d} & 
\colhead{$r_s$} & \colhead{$M_s$} & \colhead{\vm} \\
$10^{9}\msun$  &     &     &     & $\msun/$pc$^2$ &    &    & $10^{-2}$ &  kpc  & $10^8M_{\odot}$ & km/s }
\startdata
1   & V & 12 & 0.048 & 1.75 & 0.64 & 0.54 & 1.1  & 1.1 & 3.61 & 55 \\
    & Cl & 19 & 0.035 & 2.42 & 0.39 & 0.74 & 12.3 & 0.7 & 6.18 & 71 \\
100 & V & 10 & 0.048 & 2.30 & 0.50 & 0.53 & 8.8  & 6.3 & 512  & 212 \\
    & Cl & 13 & 0.035 & 2.83 & 0.27 & 0.72 & 24.5 & 3.8 & 730  & 278 \\
\enddata

\tablenotetext{a}{Environment: V=void, Cl=cluster}

\end{deluxetable}

\section{The halo-galaxy connection}

Our study shows that present-day halos in the cluster environment are on 
average more concentrated, more spherical, disaligned and rotate slower than 
halos in the void or field environments. Our study also suggests differences
in the evolution history of halos depending on the environment: halos in 
dense environments assemble their masses apparently earlier than halos
in low-density regions \citep[see also][]{ST04}. The question
is whether all these differences in the halos produce visible differences   
in the luminous galaxies formed inside them. 

A common method in galaxy modeling is to build-up a luminous galaxy linked 
directly to a present-day 
CDM halo. Several numerical and analytical works show indeed that the present-day 
properties of the halos keep some memory of their evolution history (typical 
formation epoch, average shape of the MAH, major mergers, etc.). For example, 
the $z=0$ halo concentration depends on the shape of the MAH 
\citep{AR98, FA00, Risa02,Zhao03}, in such a way that fixing the concentration 
at $z=0$, the main feature of the halo MAH (and of the corresponding baryon matter 
infall history) remains roughly determined. However, a direct ($z=0$ halo)-galaxy 
connection may fail for subhalos (specially for the massive ones and in 
cluster regions). The tumultous history of galaxy-size subhalos, since they 
infall in the cluster halo, affects in a complex way their present-day properties
and distributions. The main effect is that of tidal stripping, or even total 
disruption, of subhalos as they penetrate into the cluster potential 
\citep[\eg][]{Ghigna98, Ghigna00,Pedro99, Pedro00, Taffoni03, DeLucia04, 
Kravtsov04a, Kravtsov04b, Diemand04, Gaa04a, reed04}.
Thus, the halo mass function and the halo-to-galaxy mass ratio of galaxies in
clusters is expected to change dramatically with time. The study of galaxy
evolution in clusters, at least in their inner regions, requires a full
treatment of halo evolution and baryonic physics (Springel et al. 2001;
Diaferio et al. 2001; Gao et al. 2004b).  

Following, we will carry out a very preliminar exploration of the effects on 
the {\it disk} galaxy properties when varying the {\it present-day} CDM halo 
concentration 
and $\lambda$, according to the results obtained above for different environments. 
For this, seminumerical models of disk galaxy evolution \citep{AF00, FA00} will 
be used \citep[see also][]{vdBosch00}. These models include in a self-consistent 
way the processes of formation and evolution of a spherical CDM halo and a disk 
in centrifugal equilibrium inside it, the adiabatic gravitational contraction of 
the halo due to disk formation, self-regulated SF and feedback, secular bulge 
formation and other evolutionary processes. Most of the $z = 0$ galaxy properties 
depend mainly on (i) the present-day \mv, (ii) the halo MAH (that determines the 
concentration), (iii) the spin parameter $\lambda$, and (iv) the disk mass 
fraction $f_d$ ($\equiv M_d/\mv$). A brief description of the main 
ingredients of the model is presented in the Appendix. We notice that 
the $\lambda$ parameter in this case follows its standard definition 
\citep[\eg][]{Peebles69}. This $\lambda$ parameter is larger than the 
$\lambda'$ one measured for halos in \S 4 by factors typically of 1.25-1.50, 
depending on the halo concentration \citep{Bullock2001a}.

We model disk galaxies of a given baryon mass ($M_d$) in halos
with different concentration and spin parameters, emulating void and cluster
environments. The seminumerical method is well suited for isolated (field and
void) galaxies. For ``void'' galaxies, we assign the (high) median $\lambda$ 
(calculated from $\lambda'$) found for VOID halos and fix the
$z=0$ \mv\ ($=M_d/f_d$). We then select, from random realizations,
MAHs for this mass so that they yield the typical (low) concentrations 
of VOID halos at $z=0$; these MAHs are extended, implying late halo assembling. 
We carefully  choose  roughly regular  MAHs, without 
dramatic changes in their shapes. The emulation of ``cluster'' galaxies with 
our method is more difficult. We will assume that the (sub)halo was accreted into 
the cluster halo at $z=0.3$ (3.4 Gyr ago, for the cosmology used here). 
Ours and previous simulations show that most subhalos in present-day
cluster halos were accreted recently ($\sim 70\%$ after $z=0.5$), and that
since then, these halos have lost typically 30-50\% of their masses due to tidal
striping \citep[\eg][]{DeLucia04,Andrey05,Zentner04,vdB04}. Thus, we fix 
\mv\ at $z=0.3$
rather than at $z=0$, and we assign a low $\lambda$, typical of our CLUSTER 
halos at $z=0.3$, which is slightly larger than at $z=0$.  
Regarding concentrations, we fix them to the typical (high) concentrations of 
our CLUSTER halos by selecting the appropiate MAHs. These MAHs and the fact
that the halo is fixed at $z=0.3$, imply an early assembling of the halo/disk
system. Since  $z\lesssim 0.3$ the disk does not accretes more gas but
it continues evolving.

Models for two disk (baryonic) masses are calculated, $M_d = 10^{9}\msun$ 
and $10^{11}\msun$. For each mass, we calculate two galaxy models corresponding 
roughly to the extreme environments, void and cluster regions, as explained 
above. We set the disk mass fraction in all cases to be $f_d=0.03$. 
The cosmologycal parameters are the same ones used in our N-body simulations (\S 2). 
Table 5 summarizes the main halo and disk input parameters as well as 
the obtained properties of our simulated galaxies. Recall that in the case of the 
``cluster'' galaxies, the halo growth and gas infall are truncated at $z=0.3$ 
to account for the fact that it was accreted into the cluster at this time. 
The late-accreted halos are typically located in the periphery of the cluster halo
\citep[\eg][]{DeLucia04,Andrey05} and are expected to host mostly spiral
or S0 galaxies.

Looking at Table 5, one sees that the dependence of halo concentration 
and $\lambda$ on environment found in this work should produce some 
changes in the disk galaxy properties. As previously reported \citep{AF00}, $\lambda$ 
influences mainly the disk scale length and surface brightness, the gas fraction 
and the secular bulge-to-disk ratio, while the MAH fixes the halo concentration
and influences mainly the SF history, galaxy integral color and the scatter 
of the TFR. Thus, our expectation is that disk galaxies of a given mass $M_d$ 
formed in low--$\lambda$ and highly concentrated halos, with a gas infall 
history truncated early (cluster environment), are preferentially of
earlier morphologycal types, redder, shifted to the higher velocity side
in the TFR, and have higher surface brightness, smaller scale lengths, and lower 
gas fractions than disk galaxies formed in high--$\lambda$ and low-concentration 
halos (void environment). All these trends are namely seen in Table 5.   
The stellar central surface density, $\Sigma_0$, and disk scale radius,
$r_d$, are the parameters of the exponential-law fit to the model stellar
disk. In fact, the disks formed in CDM halos typically are not exactly 
exponentials, but their surface density profiles are more concentrated in 
the center and with an excess in the periphery compared with the exponential 
law \citep{FA00,Bullock2001a}. 

Our results show that owing to only differences in the halo properties (related
to the environment), the galaxy disks can have significant differences.
According to Table 5, disks formed in halos with properties typical of halos in
voids are $\sim 0.2$ mag bluer in $(B-V)$ color, have central stellar 
surface densities lower by $\sim 4$ times (1.5 mag/arcsec$^2$), and gas
fractions higher by 1.5-2 times than disks formed in halos with properties
typical of the cluster-periphery environment. According to our 
models, the low mass disks have lower central surface brightnesses and
higher gas fractions than the high mass ones, while the $(B-V)$ colors
do not change significantly with mass.  The (secular) bulge-to-disk
ratio depends strongly on mass, being this ratio larger for massive disks. 
There is also a significant increasing of this ratio from halos
typical of voids to those typical of the cluster periphery. This implies
that {\it part of the morphology-density relation is due to changes in the
properties of the galaxy halos with environment}.

As mentioned in \S 5.1, one expects also differences with environment in the 
TFR. The slope of the model stellar TFR is $\sim 3.4$ \citep{FA00}.
By using this slope, we correct the velocities due to the variations 
in the stellar masses of each one of the models with 
$M_d=10^{9}\msun$ and $M_d=10^{11}\msun$ presented in Table 5. 
We obtain that at $M_d=10^{11}\msun$, the difference in \vm\ for our 
void and cluster-periphery galaxies is about 20\%; the difference in 
mass, expressed in magnitudes, is 0.6 mag, i.e. we predict that 
the zero-point of the TFR of void galaxies at $M_d=10^{11}\msun$ 
should be brighter by $\sim 0.6$ mag than the one of cluster-periphery 
galaxies. At $M_d=10^{9}\msun$, the predicted difference is $\sim 0.4$ mag. 

Real galaxies show likely more pronounced differences in 
their observational properties as a function of the environment than those 
obtained with our models (see \S 1.1 for references). Thus, other 
physical processes not considered here should certainly play an important role 
in galaxy dynamics and evolution. For example, we did not treat in detail 
the angular momentum distribution misalignment and did 
not take into  account the halo triaxiality and the shape-to-rotation axis 
misalignment. These properties change with environment as was shown in \S\S 3 and 4. 
Besides, deep in the cluster regions, the external effects such as tidal and ram 
pressure stripping, strong interactions at early epochs, and galaxy harassment
\citep{Moore96}, 
are likely the dominant ones in shaping morphology and other galaxy 
properties (see the references above).  

Finally, we note that most of the differences in the halo properties
with environment seen at $z=0$ dissapear at $z=1$, although the scatters
in the latter epoch are very large. As mentioned in \S 1.1, observations
also show that the morphology-density relations flattens at higher
redshifts. The main morphological evolution is seen for the galaxies in 
the high-density environment. Halos in clusters are also those that 
change most their properties. 

\section{Summary and Conclusions}

Observations show that morphology and several properties of galaxies change 
systematically with environment. According to the current paradigm of galaxy 
formation, galaxies assemble inside CDM halos. We state then two questions: 
(i) do the properties of {\it galaxy-size} CDM halos/subhalos change systematically 
with environment?, and if this the case, (ii) do these changes 
in the halo/subhalo properties affect the luminous galaxies in the 
direction that observations show?  

We studied the first question by means of
high-resolution $\Lambda$CDM cosmological N-body simulations. Several clusters 
with their surroundings (CLUSTER sample) and void regions (VOID sample) were 
selected from large-box simulations, and they were resimulated with high 
resolution. We also use other simulations, in particular two 60\mpch-box 
simulations from which the FIELD sample of parent halos was extracted. 
The local density contrast in our CLUSTER, VOID and FIELD samples are on average 
larger than 34, smaller than -0.8 and roughly 0, respectively. The second question,
in a first attempt, was disscused by using standard seminumerical models of disk 
galaxy evolution. Disk galaxies were modeled inside CDM halos with properties
that we have found they have in the different environments. Following, we 
summarize our main results:
 
$\bullet$ For masses $\lesssim 5\times 10^{11}\msunh$, halos from the CLUSTER sample 
are on average $\sim 40\%$ more concentrated and have $\sim 2$ times higher central 
densities $\rho_{-2}$ than halos in voids at $z=0$. While for halos in cluster 
regions the concentration parameters \cnfw, \cq\ and \cd, and the density 
$\rho_{-2}$ decrease on average with mass, for halos in voids these concentrations 
and $\rho_{-2}$ do not seem to change with mass. The slope of the 
$\cnfw-\mv$, $\cq-\mv$ and $\cd-\mh$ dependences for the former sample is 
$\sim -0.1$. In the mass range of the parent FIELD halos analyzed here, 
concentrations and $\rho_{-2}$ are also smaller on average than those of halos 
in clusters. The concentrations and $\rho_{-2}$ of FIELD halos decrease on 
average with mass but less rapid than halos from the CLUSTER sample. The scatters 
of all of these parameters are larger for CLUSTER halos than for halos in less dense
environments. For example, for reasonably well fitted NFW halos from the CLUSTER
and VOID samples, $\Delta$(log$\cnfw$)$\approx 0.13$ and $\approx 0.09$,
respectively.

$\bullet$ All the differences mentioned above become less pronounced when comparing
only the parent halos from the different samples (CLUSTER: $\sim 60\%$;
VOID: $\sim 95\%$; FIELD: all). Therefore, the CLUSTER halos are more concentrated
and internally denser than the VOID and FIELD ones due partially to a local
halo-subhalo effect (subhalos are a significant fraction of the CLUSTER sample,
and subhalos are systematically more concentrated than their parent halos,
see also, e.g., Bullock et al. 2001a). However, we find that the parent halos 
and subhalos from the CLUSTER sample are still significanlty different on 
average than the parent halos and subhalos from the VOID and FIELD samples. 
Therefore, the differences in halo properties are certainly also
due to a pure global environmental effect, related mainly to the typical
halo formation epoch: {\it halos in dense environments assemble their masses
earlier than halos in low-density regions}.
 
$\bullet$ Halos in dense environments are more spherical than halos
in less dense environments at $z=0$. The minor-to-major axis ratios 
of CLUSTER halos are on average $\sim 1.2$ times lower than those of the FIELD
halos. For a given epoch, the ellipticity of the halos tends to 
increase with mass. For a given mass, the ellipticity changes with age,
the younger halos having on average larger ellipticities than the older halos.

$\bullet$ The spin parameter of CLUSTER halos is on average $1.3-1.4$ times 
lower than the one of VOID or FIELD halos at $z=0$. This is likely a consequence 
of both global (environmental) and local effects, i.e. the
dependence of halo formation epoch on environment, and the tidal stripping 
and ``tumultuos'' histories that halos suffer in locally high-density
regions, respectively. We do not find significant differences in the 
${\lambda'}$ distribution of parent halos and subhalos from the CLUSTER sample.  
The CLUSTER halos appear to have a less aligned intrinsic angular momentum 
distribution than the VOID and FIELD ones, but this result needs to be 
confirmed by future analysis because of the large errors involved in the 
determination of \costha. The spin parameter does not change significantly with 
mass in any environment and
its distribution is well approximated by a lognormal function with a
larger width and a lower peak ${\lambda'}$ in cluster regions
than in voids and the field.  The angular momentum axis of halos tends to be 
aligned with their minor principal axis, this behaviour being more common 
in the less dense environments than in the high-density ones. 

$\bullet$ A tight $\vm-\mh$ relation is seen for halos in all the 
environments. The relation is shallower (slope of 0.30) and more scattered 
for clusters and their surroundings than for the void and field environments
(slopes of $\sim 0.33-0.34$). If this
relation is the basis of the observed TFR,
then slight differences in the TFR are expected in different environments.
The $\vm-\mh$ relation slightly shifts to the lower mass side at $z=1$
in all the environments, but in particular in the voids.  Similar results
to the $\vm-\mh$ relation are obtained for the $\sigc-$ and $\sigt-\mh$
relations, showing that most of the halos are close to their equilibrium
state.  

$\bullet$ The differences in halo properties with environment seen at $z=0$ 
drastically diminish at $z=1$. Interestingly enough, a similar result
was found for the observed morphology-environment relation. As expected, 
the concentration parameters decrease as we go from $z=0$ to $z=1$, being 
this change more pronounced for the CLUSTER halos, particularly 
for the less massive ones. The spin 
parameter of CLUSTER halos at $z=1$ is on average significantly 
larger than at $z=0$, while for less dense environments, the 
spin parameter at $z=1$ is on average only slightly larger
than at $z=0$; thus, $\lambda'$ evolves in a different way for halos
in cluster-like regions than for halos in less dense environments. Halos 
in the void and field environments are systematically 
less aligned at $z=1$ than at $z=0$, while halos in the cluster
regions do not show any systematical change with redshift in a 
statistical sense.  
 
$\bullet$ Disk galaxies modeled in a self-consistent fashion 
inside $\Lambda$CDM halos with the present-day concentrations
and spin parameters found here for halos in the different environments
present systematical differences: the galaxies formed in halos typical 
of cluster-periphery environment have higher surface density, circular 
velocity and secular bulge-to-disk ratio, lower gas fraction, and are redder 
than for those formed in halos typical of void environment. These trends
agree qualitatively with observations but are not enough to explain the 
observed differences of galaxy properties with environment. We predict
that the TFR of galaxies in low and high-density environment is different.

From our study we conclude that most of the properties of galaxy-size 
halos at $z=0$ change with environment in a statistical sense, the 
largest differences in the structural properties being for the less 
massive halos (sub-$L_*$ galaxies). The main changes with 
environment occured after $z\sim 1$ and the most affected halos 
are the subhalos in the CLUSTER sample. The differences in halo concentrations 
and spin parameters along the different environments influence
on the properties of galaxy disks formed inside these halos and 
in the same direction that observations show. However, the inclusion of 
angular momentum dissalignment, triaxiality, and shape-rotation
axis dissalingment --halo properties that also change with environment--
in the models of disk galaxy evolution is necessary in order to attempt 
to reproduce in more detail the observed relations of disk galaxy properties 
with environment. The halo merging history and its influence
on luminous galaxies should also be taken into account, in particular
for modelling galaxies in the high-density environments. 

The halo properties discussed above and their changes with environmnet 
are ultimately related to the {\it initial cosmological conditions}. 
Based on our results, we claim that the observed galaxy properties--environment 
relations are partially established by the initial cosmological 
conditions (nature), in the sense that galaxy halos in more dense environments
assemble earlier. However, our results point out that the {\it external astrophysical 
factors} (nurture) should play also an important role in the observed trends 
of changing galaxy properties with environment, specially in the highest density 
regions (clusters).

\acknowledgments

Computer simulations presented in this paper were done at the Leibnizrechenzentrum 
(LRZ) in Munich and at the John von Neumann Institute for Computing J\"ulich.
We acknowledge the anonymous referee whose helpful comments and suggestions 
improved several aspects of the paper.
We are also grateful to Ricardo Flores for kindly providing  
a copy of his program to compute the halo ellipticities. 
This work has been supported by a bilateral CONACyT-DFG (Mexico-Germany) grant,
and by CONACyT grants 36584-E and 40096-F.

\vspace{.9cm}

\centerline{\bf APPENDIX} 

\vspace{0.4cm}

Here we present the main physical ingredients of the self-consistent 
evolutionary models used in \S 6. For details see \citet{FA00,AF00}.
The disk is built up within a growing $\Lambda$CDM halo. An extended
Press-Schechter approach \citep{LC93,AR98} is used to generate the statistical 
MAHs of the halos from the primordial density fluctuation field, 
and a generalized secondary infall model is applied to calculate the 
time by time virialization of the accreting mass shells \citep{AR98}.
The evolution and structure of the $\Lambda$CDM halos calculated this 
way agree well with results from cosmological N-body simulations \citep{AR99}. 
Halos assembled through early active MAHs end more concentrated on average
than halos with extended MAHs. 

The halo mass shells are assumed to have aligned rotation axis with specific 
angular momentum given by $j_{sh}(t_v)=dJ(t_v)/dM_v(t_v)$,  where
$J=\lambda GM_v^{5/2}/\left| E\right| ^{1/2}$, $J$, M$_v$ and $E$ are
the total angular momentum, mass and energy of the halo at the shell
virialization time $t_v$. The spin parameter, $\lambda$, is assumed to
be constant in time. As the result of the assembling of these
mass shells, a present day halo ends with an angular momentum distribution
close to the (universal) distribution measured by \citet{Bullock2001a} 
in N-body simulations. A fraction $f_d$ of the mass of each shell is assumed
to cool down and form a disk layer in a dynamical time. The radial mass 
distribution of the layer is calculated by equating its specific angular
momentum to that of its final circular orbit in centrifugal equilibrium
(detailed angular momentum conservation is assumed). The superposition
of these layers form the disk. 

The gravitational interaction of disk and inner halo is calculated using a 
generalized adiabatic invariance formalism. This formalism differs from the usual 
one \citep[e.g.,][]{Flores93} in that we take into account the ellipticity of the 
orbits (the circular orbit assumption is relaxed); recent numerical studies 
\citep{Gnedin04} confirm the validity of our approach. 

The disk SF at a given radius (azhimutal symmetry is assumed) is triggered by 
the Toomre gas gravitational instability criterion and self-regulated by a 
vertical disk balance between the energy input due to SNe and the turbulent 
energy dissipation in the ISM. This physical prescription for disk SF 
yields naturally a Schmidt law with an index $n\lesssim 2$, slightly 
varying along the disk \citep{FHG96}. The SF efficiency depends on 
the gas surface density determined mainly by $\lambda$, and on the gas accretion
rate determined by the cosmological MAH. Finally, we estimate the mass
of the (secularly formed) bulge as the inner disk mass where the Toomre 
stellar parameter indicates disk instability  \citep[see also][]{vdB98}

\end{document}